\documentclass[journal=jacsat,manuscript=article]{achemso}

\usepackage[version=3]{mhchem} 
\usepackage{subfiles}
\usepackage{amsfonts}
\usepackage{amsmath}
\usepackage{graphicx}
\usepackage{subcaption}
\usepackage{hyperref}
\usepackage{xcolor}
\usepackage{float}

\newcommand{\Vx}{\mathbf{x}}

\newcommand{\VH}{\mathbb{H}}

\newcommand{\VD}{\mathcal{D}}
\newcommand{\VX}{\mathcal{X}}
\newcommand{\VY}{\mathcal{Y}}

\author{Jize Zhang}
\affiliation{Center for Applied Scientific Computing, Computing Directorate, Lawrence Livermore National Laboratory}
\email{zhang64@llnl.gov}
\author{Bhavya Kailkhura}
\affiliation{Center for Applied Scientific Computing, Computing Directorate, Lawrence Livermore National Laboratory}
\email{kailkhura1@llnl.gov}
\author{T. Yong-Jin Han}
\affiliation{Materials Science Division, Physical and Life Sciences Directorate, Lawrence Livermore National Laboratory}
\email{han5@llnl.gov}

\title[Reliable Materials Informatics]
  {Leveraging Uncertainty from Deep Learning for Trustworthy Materials Discovery Workflows}

\abbreviations{IR,NMR,UV}

\begin{document}


\begin{abstract}
In this paper, we leverage predictive uncertainty of deep neural networks to answer challenging questions material scientists usually encounter in machine learning based materials applications workflows. First, we show that by leveraging predictive uncertainty, a user can determine the required training data set size necessary to achieve a certain classification accuracy. Next, we propose uncertainty guided decision referral to detect and refrain from making decisions on confusing samples. Finally, we show that predictive uncertainty can also be used to detect out-of-distribution test samples. We find that this scheme is accurate enough to detect a wide range of real-world shifts in data, \emph{e.g.,} changes in the image acquisition conditions or changes in the synthesis conditions. Using  microstructure information from scanning electron microscope (SEM) images as an example use case, we show that leveraging uncertainty-aware deep learning can significantly improve the performance and dependability of classification models.
\end{abstract}

\section{Introduction}

Deep Learning (DL) techniques are achieving remarkable success in a wide range of scientific applications~\cite{raghu2020survey}. 
One such application is Materials Informatics that applies DL methods to accelerate the discovery, synthesis, optimization and deployment of materials~\cite{ramprasad2017machine}. 
As a motivating example, consider the problem of \emph{Materials Discovery} where one is interested in screening novel materials that meet certain performance requirements~\cite{kailkhura2019reliable}. A critical obstacle in developing and deploying materials in a timely manner is understanding the complex process, structure, property and performance (PSPP) relationships, including evaluating the property of interest of a material, which can take a significant amount of time and effort. 
DL offers opportunities to potentially accelerate this process by learning the complex structure–property relationships (e.g., materials' physical, mechanical, optoelectronic, and thermal properties), from historical characterization and property data (i.e., training data) and produce models that can predict material properties at dramatically lower overall costs on unseen test data. Thus, DL is becoming increasingly prevalent in Materials Informatics workflows for making potentially important decisions, thereby making applications of DL approaches in Materials Discovery application a high-stake in nature. For example, incorrect decisions or predictions in Material Discovery can have (a) significant costs in terms of experimental resources and time when testing bad performing materials recommended by DL, or (b) lost opportunities to discover high performing materials rejected by DL. 

\begin{figure}[!t]
\centering
    \includegraphics[width=\textwidth]{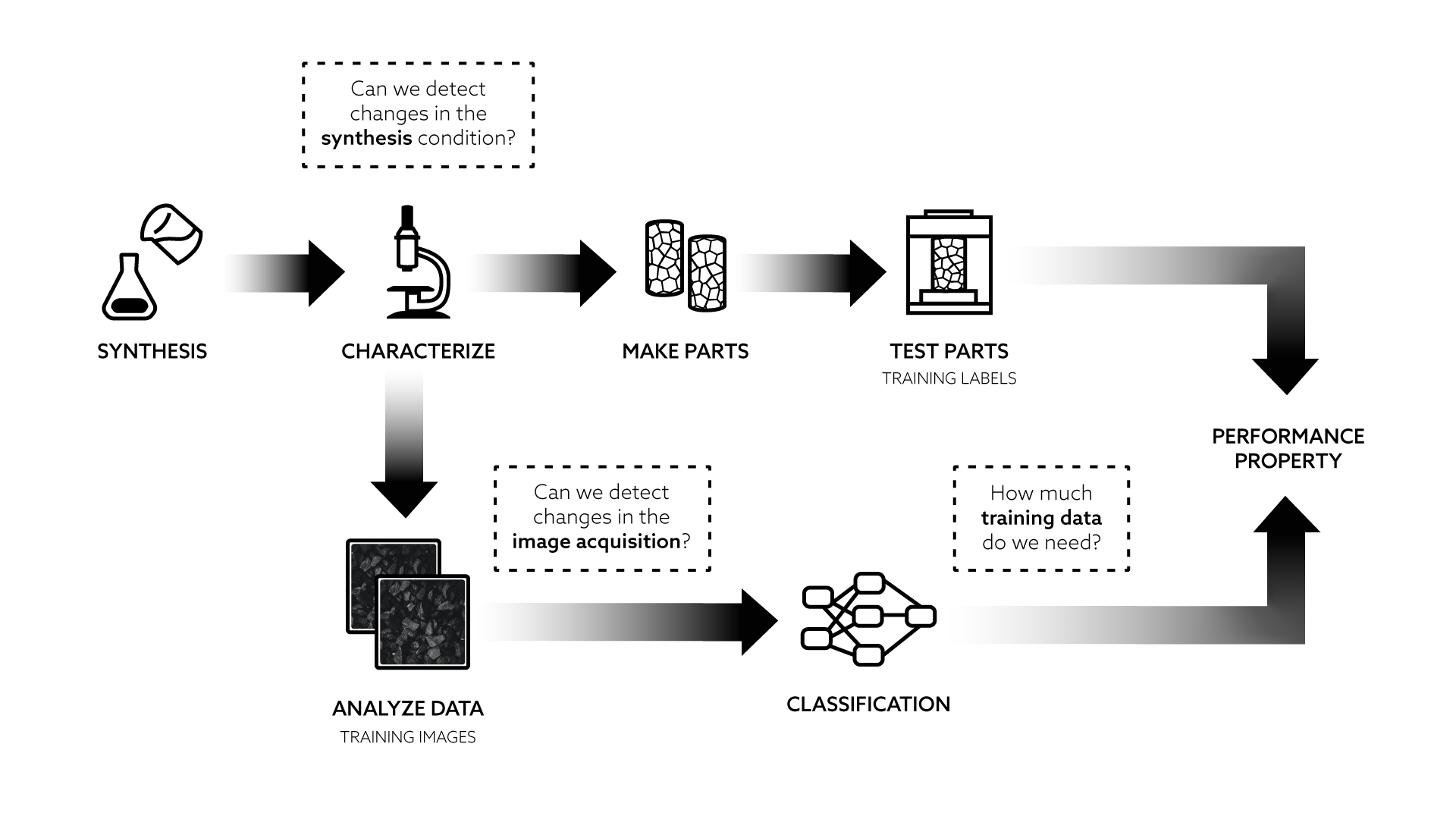}
    \caption{A schematic of a workflow for SEM image based material classification task.}
    \label{fig:MI}
\end{figure}

The majority of efforts in Materials Informatics are currently devoted to training deep neural networks (DNNs) that can achieve high accuracy on holdout dataset from the training distribution~\cite{gallagher2020predicting,ling2017building}. 
Existing efforts implicitly assume ideal conditions during both training and testing by assuming (a) having access to a sufficiently large labelled training data and (b) test data from the ``same distribution" as the training set. Unfortunately, these conditions are seldomly met in Materials Discovery applications since a large amount of \emph{relevant} historical data is rarely available, and the test data is typically and systematically different from the training data either through noise{\color {black} \cite{liu2017materials}} or other changes in the distribution{\color {black} \cite{haghighatlari2019thinking}}. This is an inherent challenge in applying DL to scientific domain whose aim is to find something ``new" and ``different" that outperforms existing materials. It is well known that DL models are highly susceptible to such distributional shifts, which often leads to unintended and potentially harmful behavior (i.e., overconfident on predictions) especially when trained with insufficient amount of data~\cite{Bulusu20}. 
Therefore, to ensure that the trained models behave reliably and to establish DL as a dependable solution in high-stake Materials Discovery applications, the following equally important questions needs to be answered.

\noindent\textbf{1. How much data is required to train a DNN?}

The quality and amount of training data is often the most important factor that determines the accuracy of a DL model. However, determining the amount of data necessary to train a DNN to achieve high classification accuracy is a challenging problem. A reasonable approach is to model validation accuracy as a function of the amount of training data to predict the optimal data size needed to achieve a certain classification performance (referred to as learning curve based approach)~\cite{cho2015much}. Unfortunately, to get a reliable estimate of validation accuracy, this approach requires a large amount of labelled data not used in the training process. 
This is clearly not feasible for the application of interest where labelling process (e.g., performing experiments to collect relevant data) is expensive and time consuming.

To overcome this challenge, we propose to leverage the DNN prediction uncertainty (or confidence) information and approximate the validation accuracy in the learning curve using unlabeled data only. Specifically, we obtained the average confidence (i.e., the predicted probability of the most plausible label) over the set of unlabeled data as the estimated validation accuracy for generating the learning curve. We show that having access to unlabeled scanning electron microscopy (SEM) images is sufficient in our approach to determine with high precision how many more material samples should be imaged and collected to provide sufficient DL training data to achieve {\color {black} the desired accuracy label}.

\noindent\textbf{2. How to equip DNNs with a reject option?}

The performance and dependability of DNN {\color {black} classification models} can be dramatically improved by building in a reject option. Testing the DL model on difficult (or ambiguous) material samples is a common case in material science applications since the discovery of new materials often involve moving away from existing materials using different synthesis and processing conditions. In such a context, having a reject option allows DNN models to refer a subset of difficult material samples for further evaluation and testing while making predictions on the rest. This is also known as selective classification where the classifier abstains from making a decision when the model is not confident, while keeping coverage as high as possible. An added benefit to deploying selective classification is that this approach has the potential to result in substantial improvements in the performance on the remaining materials data. Unfortunately, many off-the-shelf ML models are not well calibrated, i.e., the probability associated with the predicted class label does not match the probability of such prediction being correct {\color {black} \cite{niculescu2005predicting, guo2017calibration}}. It has been observed that this issue is more noticeable in complex models such as DNNs~\cite{zhang2020mix}. Intuitively, this observation implies that DNNs are particularly bad at recognizing ambiguous samples.

To overcome this challenge, we propose to use state-of-the-art uncertainty-aware DNNs {\color {black} \cite{fort2019deep,ovadia2019can}} in Materials Discovery workflows as they are known to be better calibrated as compared to their baseline counterparts. Specifically, we show that uncertainty guided decision referral can substantially improve the classification accuracy on the non-referred samples while reducing the number of referred (i.e., rejected) samples.

\noindent\textbf{3. How to make DNNs recognize Out-of-Distribution examples?}

Test data in real-world DL applications usually deviates from the training data distribution, e.g., due to sample selection bias, non-stationarity, and noise corruptions in some cases~\cite{Bulusu20}. For example, in the material science application (i.e., analysis of SEM images), the deviation between train and test data may arise due to: (a) change in the equipment used to capture the training material samples, or (b) changes in the material synthesis process. Having plentiful training data that can potentially cover all possibilities and variance is ideal, but in practice, that is rarely attainable. Thus, a desirable feature in Materials Discovery workflow is for a model to be aware and not be very confident on test data that is very far (or different) from the data used to train it. For example, a potentially novel material that is different from all training data should result in a request for expert help rather than misclassification into a known material class as the detection of undiscovered material is in fact a task of significant interest.
In other words, we need to require DNNs to be aware of cases when data acquisition setup or synthesis conditions are so different from the ones used during training that DL predictions cannot be trusted. Unfortunately, DNNs often make overconfident misclassification in the presence of distributional shifts and Out-of-Distribution (OOD) data~{\color{black}\cite{hein2019relu}}. Accurate predictive uncertainty is a highly desirable property in such cases as it can help practitioners to assess the true performance and risks to decide whether the model predictions should (or should not) be trusted~\cite{mallick2020probabilistic}.

Similar to the second question, we propose to leverage the predictive uncertainties produced by the state-of-the-art uncertainty-aware DNNs to overcome this challenge in Materials Discovery workflows. Specifically, we show that predictive uncertainty of uncertainty-aware DNNs are indicative enough to differentiate among the in-distribution data, data captured with different equipment, and data generated with different synthesis conditions.

To put these questions into context, we consider the problem of differentiating materials based on their microstructures as observed by their SEM images. This problem is formulated as an \emph{M}-class classification problem where each class corresponds to a specific material, with unique characteristics determined by a unique set of synthesis parameters.

We use deep neural networks (DNNs) to determine whether DL models can learn to differentiate materials purely based on their complex microstructures, which are often challenging by human assessment for samples with similar looking microstructures but very different mechanical test behaviors (\autoref{fig:MI}). By determining the accuracy of the DL models along with prediction uncertainties, we may begin to understand what microstructures lead to samples with desired performance metrics and avoid extensive testing and evaluation for \emph{newly synthesized} materials with high confidence predictions. By demonstrating that DL can featurize microstructural information from SEM images and classify each material with high accuracy, we can start to build confidence in DL models to screen, optimize, and discover materials in significantly shorter amount of time compared to an Edisonian approach.

\subsection{Main Contributions}

Deep Learning based solutions for Materials Informatics applications have so far been suggested some times without considering these important questions listed in the previous section . Yet, techniques to ensure the dependability of automated decisions are crucial for integrating DL in Materials Discovery workflows. The main contribution of this paper is to show that uncertainty-aware DL is a unified solution that is capable of answering all these questions by leveraging predictive uncertainty of DNNs. 
We demonstrate the applicability of our technique by using DL models to classify the microstructural differences of a material based on their corresponding SEM images.
Summary of our findings are as follows.
\begin{itemize}
    \item First, we show that by leveraging predictive uncertainty one can estimate classification accuracy at a given training sample size without relying on labelled data.
    This serves a general methodology for determining the training data set size necessary to achieve a certain target classification accuracy.
    \item Next, we show that predictive uncertainty can be a guiding principle to decide which material samples should be referred to a material scientist for further testing and evaluation instead of making a DL based prediction. We find that this uncertainty guided decision referral can dramatically improve the classification accuracy on the remaining (i.e., non-referred) examples.
    \item Finally, we show that predictive uncertainty can be used to detect distributional changes in the test data. We find that this scheme is accurate enough to detect a wide range of real-world shifts, e.g., due to changes in the imaging instrument or changes in the synthesis conditions.
\end{itemize}

Although we focus on a specific materials application in this paper, the proposed methodology is quite generic and can be used to make the application of DL to a wide range of scientific domains dependable and trustworthy. 



\section{Methods}

\subsection{Data Sets}
\begin{figure}[!t]
\centering
    \includegraphics[width=.5\textwidth]{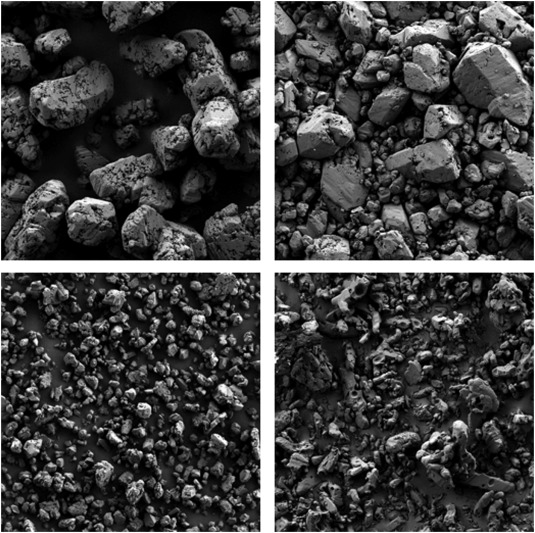}
    \caption{ Representative SEM images to illustrate the typical microstructural variability for different TATB lots. The varying particle size, porosity, polydispersity, and facetness can be clearly observed. Images have been processed to normalize image contrast and brightness levels. {\color {black} Reprinted with permission from Ref.~\citenum{gallagher2020predicting} (CC BY 4.0)}.}
    \label{tatb}
\end{figure}

Our main dataset is comprised of SEM images of 30 different lots of 2,4,6-triamino-1,3,5-trinitrobenzene (TATB). Here, a lot refers to TATB crystals produced under a specified synthesis/processing condition. TATB is an insensitive high explosive compound of interest for both Department of Energy and Department of Defense \cite{willey2006changes}. After each lot has been synthesized (with different synthesis conditions), each lot is analyzed with a Zeiss Sigma HD VP SEM to produce high-resolution scanned images, while holding image acquisition conditions (e.g., brightness and contrast settings) fixed across all lots/images.  Each image tile consists of 1000$\times$1000 pixels, with a corresponding field of view of 256.19 $\mu$m$\times$ 256.19 $\mu$m.  The combined images captured for the 30 lots resulted in 59,690 greyscale SEM images. Thus, labelled data of interest for DNNs corresponds to 59,690 greyscale SEM images labeled with unique designators per class (30 in total). Example SEM images for TATB for some classes are provided in \autoref{tatb}. One can notice strong visual discrepancy across SEM images from different lots (or classes).

\subsection{Deep Learning Models}

Let $\VX$ denote the SEM images and $\VY=\{1,\ldots,K\}$ represents the $K$ classes of materials. We use $\VD=\{\Vx^{(i)},y^{(i)}\}_{i=1}^{N}$ to represent the $N$ training data points (pairs of images and labels). {\color {black} Before being fed into the model, the SEM images will be down-sampled to the resolution level of 64 by 64, and the greyscale on each pixel will be normalized into the range of [0, 1].}
Given training and validation dataset, our goal is then to learn a classifier that predicts the quality of material samples in unseen test dataset. 

We trained the following vanilla and uncertainty-aware models: 
\begin{itemize}
    \item \textit{Vanilla Softmax}: Our uncertainty-unaware baseline simply regards the softmax outputs provided by the DNN as the predictive probabilities. Unfortunately, high confidence softmax predictions can be woefully incorrect, and may fail to indicate when they are likely mistaken\cite{guo2017calibration}, or detecting OOD \cite{Bulusu20}.
    \item \textit{Dropout}: We use Dropout \cite{gal2016dropout}, a variational-inference based Bayesian uncertainty quantification approach. During the training process, Dropout DNN is trained to minimize the approximating distribution (i.e., a product of the Bernoulli distribution across the DNN parameters) and the Bayesian posterior for the DNN parameters. At inference time, Dropout predicts the outputs by Monte-Carlo sampling the network with randomly dropped out units and averaging them, which is equivalent to integrating the posterior distribution and the predictive likelihood. Simple and computationally lightweight nature of Dropout may provide approximated posteriors that are inaccurate in some scenarios \cite{louizos2017multiplicative,kuleshov2018accurate,cortes2019reliable}. 
    \item \textit{Deep Ensemble}: Finally, we include Deep Ensembles \cite{lakshminarayanan2017simple}, a practical, scalable and non-Bayesian uncertainty quantification alternative for DNNs. As the name suggests, the core idea is to train the DNN classifier in the identical manner (i.e., same model architecture, training data and training procedure) multiple times, but each time with a random initialization of model parameters. With $T$ DNN classifiers (parameterized by $\theta^t,t=1,\ldots,T$) being included in the ensemble, the prediction probability vector for Deep Ensembles is the averaged softmax vector of each DNN.
\end{itemize}

\begin{figure}[!t]
\centering
    \includegraphics[width=.75\textwidth]{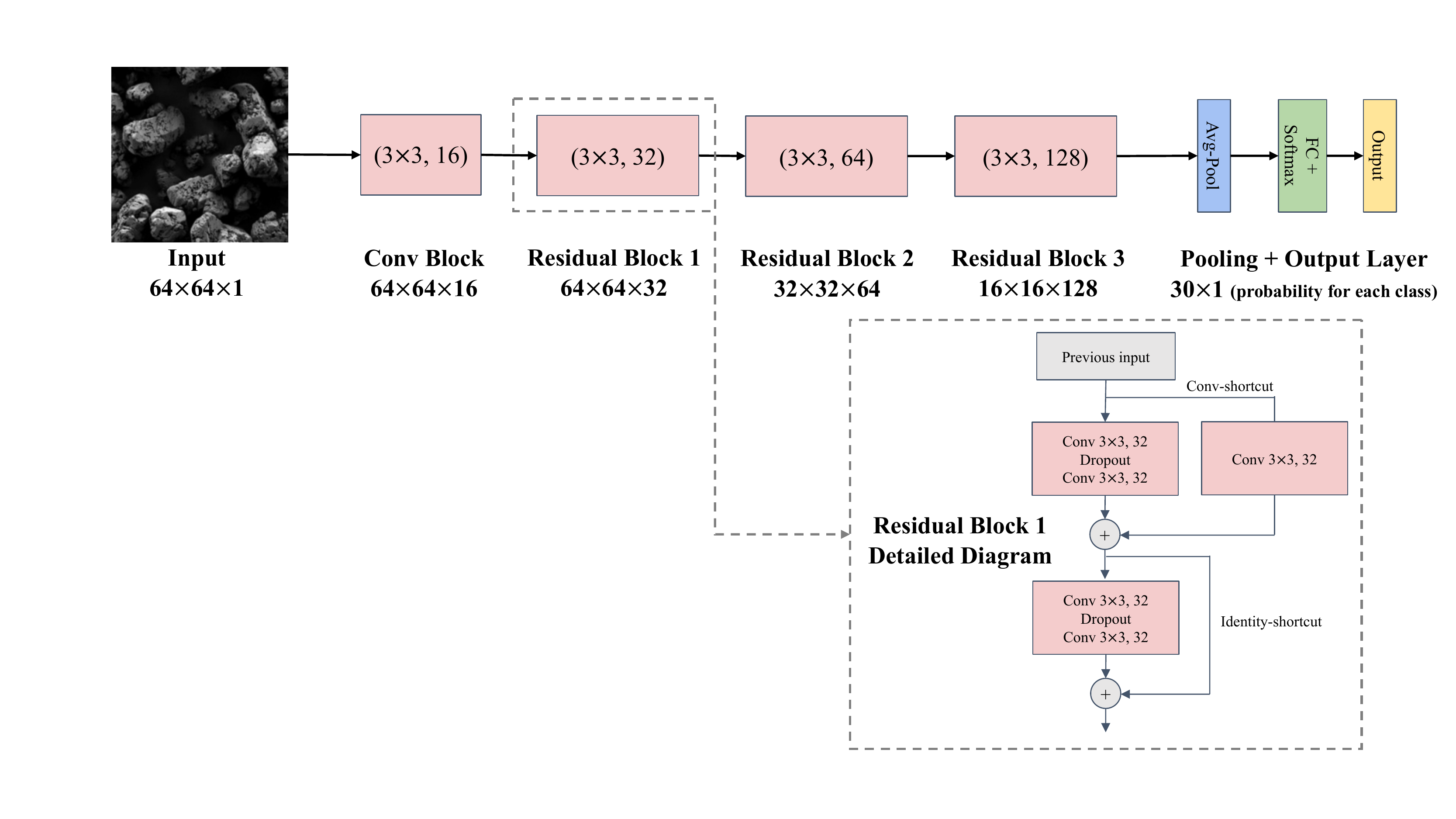}
    \caption{ A Wide ResNet architecture with a depth of 16 and a width of 2. The notation (k$\times$k, n) in the convolutional block and residual blocks denotes a filter of size k and n channels. The dimensionality of outputs from each block is also annotated. The detailed structure of the residual block is shown in the dash line box. Note that batch normalization and ReLU precede the convolution layers and fully connected layer but omitted in the figure for clarity.}
    \label{wrn}
\end{figure}

We use a Wide Residual Network (WRN) \cite{zagoruyko2016wide} architecture due to its strong performance on benchmark computer vision datasets \cite{rawat2017deep}. 
{\color {black} We apply a depth of 16 and a widen factor of 2, and the summary network structure is given in \autoref{wrn}.} Given WRN architecture, we train these DNN models with the cross entropy loss function and the Adam optimizer \cite{kingma2015adam}. The DNN model is trained from scratch (therefore, no pre-training). We set the learning rate of 0.001, and decay the learning rate by half every 50 epochs. We use a minibatch size of 64 and a weight decay factor of $5e^{-4}$. {\color {black} Hyperparameters (including learning rate, weight decay factor, as well as the network depth and width) were determined through the HpBandster toolbox, an efficient tool for hyperparameter optimization \cite{falkner2018bohb}. }
We used 200 epochs with early stopping mechanism to terminate training when the validation performance did not improve after 100 epochs. For Dropout, we used a dropout rate of $p=0.3$ and $T=16$ dropout samples for inference. For Deep Ensembles, we train $T=16$ models. All other hyperparameters are kept the same as in the standard baseline case. 
No image pre-processing technique was adopted in the training process. Horizontal flips were used for training data augmentation. We randomly divide the labelled SEM image dataset into 80\%, 10\%, and 10\% splits for training, validation and testing, respectively. {\color {black}The codes can be found online \footnote{Codebase: \url{https://github.com/zhang64-llnl/Uncertainty-DL-Material-Discovery}.}.} 

\subsection{Performance Evaluation Metrics}
We evaluate the performance of DNN models in terms of their (a) predictive performance, and (b) predictive uncertainty quality {\color {black} on the testing dataset}. 

The predictive performance is measured using the \textbf{Classification Accuracy} metric (i.e., the percentage of correct predictions among all data points).
On the other hand, following ref.~\citenum{gal2015thesis}, we use the Shannon-entropy \cite{shannon1948mathematical} as the metric to quantify the uncertainty inside the prediction probability vector $p(y|\Vx,\VD)=[p(y=1|\Vx,\VD),\ldots,p(y=M|\Vx,\VD)]$:
\begin{equation*}
\VH(y|\Vx,\VD) := -\sum_{k=1}^{K} p(y=k|\Vx,\VD)\log p(y=k|\Vx,\VD),
\end{equation*}
Basically, it captures the average amount of information contained in the prediction: $\VH$ attains its maximum value when the classifier prediction is purely uninformative (assign all classes equal probability $1/K$), and attains its minimum value when the classifier is absolutely certain about its prediction (assign zero probability to all but one class).
The quality of the predictive uncertainties is quantified using the following metrics:

\begin{itemize}
    \item \textit{Negative Log-likelihood (NLL):} It is a standard metric of the uncertainty quality by calculating the log of the joint probabilities of predictions on all test samples \cite{gneiting2007strictly}: 
\begin{equation*}
    NLL = \frac{1}{N}\sum_{i=1}^N -\log p(y=y^{(i)}|\Vx^{(i)},\VD)
\end{equation*}
Lower NLL indicates better uncertainty quality.
    \item \textit{Expected calibration error (ECE):} Calibration accounts for the degree of consistency between the predictive probabilities and the empirical accuracy. We adopt a popular calibration metric ECE \cite{naeini2015obtaining}, measuring the average absolute discrepancies between the prediction confidence versus the accuracy:
\begin{equation*}
    ECE= \frac{1}{N_b}\sum_j^{N_b} \frac{B_j}{N} |acc(B_j)-conf(B_j)|
\end{equation*}
where we sort the test data points according to their confidence (the prediction probability for the most likely label, i.e., $\max_y p(y|\bf{x})$), and bin them into $N_b$ quantiles. Here $acc(B_j)$ and $conf(B_j)$ is the average accuracy and confidence of points in the $j$th bin, and $B_j$ is the number of data points in such bin. We use 20 equal-spaced bins to measure ECE in this paper. Lower ECE is more favorable.
\end{itemize}

\section{Results and Discussion}
\label{sec:results}

In this section,  we first provide details on the DL models and their performance on the in-distribution test data. Then, we discuss and evaluate the performance of proposed solutions on the three aforementioned use cases.

\subsection{Performance Evaluation Result}

{\color {black} On the test dataset}, \autoref{metrics} shows that Deep Ensembles outperforms rest of the methods in Accuracy and ECE. Dropout achieves the best NLL and also a much better ECE than the Softmax baseline. Intriguingly, the two uncertainty quality metric ranks Deep Ensembles differently (best ECE and worst NLL). This might be attributed to the well-known pitfall of NLL to over-penalize the existence of samples with very low prediction probabilities for their true classes\cite{ovadia2019can}. Thus, we recommend ECE as the default uncertainty quality metric to avoid such pitfall and also for its better interpretability (accuracy vs confidence). Overall, these results demonstrate the effectiveness of uncertainty-aware DNN approaches over the Softmax baseline. It is important to remember that their performance gain over the baseline is not free. In the case of Deep Ensembles, additional cost must be spent on training and inference. For example, we trained an ensemble of 16 classifiers independently, and the computational cost is 16 times higher than the baseline. For Dropout, albeit the training cost is comparable to baseline, the inference computation is similarly expensive as Deep Ensembles. 

The success of Deep Ensembles is anticipated. After all, ensembling multiple machine learning models is a well known treatment to reduce the error of a single model, which has been theoretically \cite{dietterich2000ensemble,shi2018crowd} and empirically \cite{fernandez2014we,cortes2018deep} explained. Comparing to Dropout, the other ensemble-like technique, we hypothesize that Deep Ensembles learns multiple models whose predictions are much more diverse (lowly correlated) given the high-dimensionality and nonconvexity of DNN parameter spaces, which is crucial for enhancing the classification error and the uncertainty quantification quality \cite{kendall2017uncertainties}.

\begin{table}[!t]
  \caption{Accuracy and uncertainty quality of different methods.}
  \label{tbl:example}
  \begin{tabular}{lllll}
    \hline
    Approaches  & Accuracy ($\uparrow$) & NLL ($\downarrow$) & ECE ($\downarrow$)  \\
    \hline
    Baseline  & 92.3\% & 0.919 &  5.38\% \\
    Dropout & 92.1\% &  \textbf{0.903} & 2.79\% \\
    Deep Ensembles  & \textbf{95.3\%} & 0.920  & \textbf{1.52\%} \\ 
    \hline
  \end{tabular}
  \label{metrics}
\end{table}

\subsection{DL Trustworthiness Case Studies}
In this section, we show how uncertainty scores can be leveraged to answer the aforementioned important problems to make the Materials Discovery workflows dependable.

\subsubsection{Case Study 1: How much data is required to train a DNN?}
The need for large amounts of labelled training data is often the bottleneck to the successful deployment of machine learning models. This is especially crucial for DNNs due to their over-parameterized nature \cite{krizhevsky2012imagenet,lecun2015deep}. Yet, in scientific applications, obtaining high-fidelity labels can be expensive due to the associated costly and time consuming experiments \cite{schmidt2019recent}. Therefore, an important issue is to decide how much training data is required to achieve the desired accuracy level, which allows the user to prioritize experimental plans accordingly. 

Conventionally, this task is done by generating the learning curve \cite{cho2015much}, which approximately represents the relationship between the training data amount and the validation accuracy on a set of labelled data unused in training \cite{krizhevsky2012imagenet}. One can even further predict the needed training data size to achieve the required accuracy by extrapolating the learning curves \cite{domhan2015speeding,kolachina2012prediction}. However, a drawback of such conventional validation-based learning curve approach is its reliance on a large amount of labelled data unused in training to accurately evaluate the validation accuracy, which can be expensive or even infeasible in many applications. Here, we ask the question whether we can leverage the predictive uncertainty information to solve this use case with access to unlabeled validation data (i.e., SEM images without material class labels) only. Specifically, we decide to test if the average prediction confidence (which can be computed without any label information) on the unlabeled dataset can be used as a surrogate to assess the validation accuracy and approximately generate the learning curve. The logic behind such approach is that as we have discussed in the previous sub-section, for DNN models with well-calibrated uncertainties (low ECE), the average confidence should closely match the accuracy.

\begin{figure}[!t]
\centering
    \includegraphics[width=\textwidth]{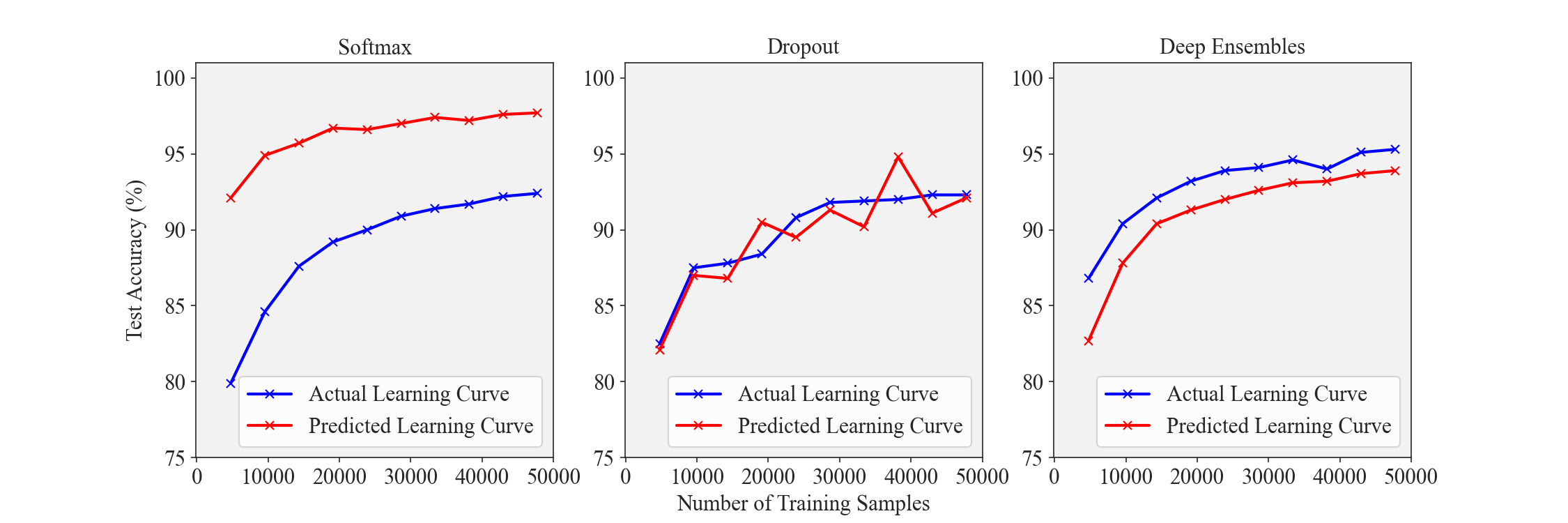}
\caption{\textbf{Uncertainty-guided learning curves.} The predicted (confidence-based) and actual learning curves using different approaches.}
\label{curve}
\end{figure}

To examine the feasibility, we train DNN classifiers with varying amount of training data (ranging from $10\%$, $20\%$ to $100\%$ of the maximum available training dataset size). We monitor the average validation accuracy as well as the predicted accuracy based on confidence, and plot the corresponding learning curves at \autoref{curve}. We see a significant difference between curves for Softmax. Specifically, it over-estimates the validation accuracy and will result in under-estimating the needed training data amount. For Dropout, the two curves consistently stay close, but seem to be weakly correlated, since the average confidence can sometimes decrease while the validation accuracy keeps improving. This weak correlation can be harmful in certain scenarios. For example, the users might want to determine if the DNN performance continues to improve as training data grows, and the Dropout predicted learning curves may lead them into wrong decisions. For Deep Ensembles, the predicted learning curve not only closely matches the actual one, but also showed nearly identical trends. 

\textbf{{To summarize, these results show that uncertainty scores from both Dropout and Deep Ensembles can be leveraged to predict the required amount of training data to achieve a certain validation accuracy without having access to labelled validation data.}}

\subsubsection{Case Study 2: How to equip DNNs with a reject option?}

\begin{figure}[!t]
\centering
    \includegraphics[width=0.6\textwidth]{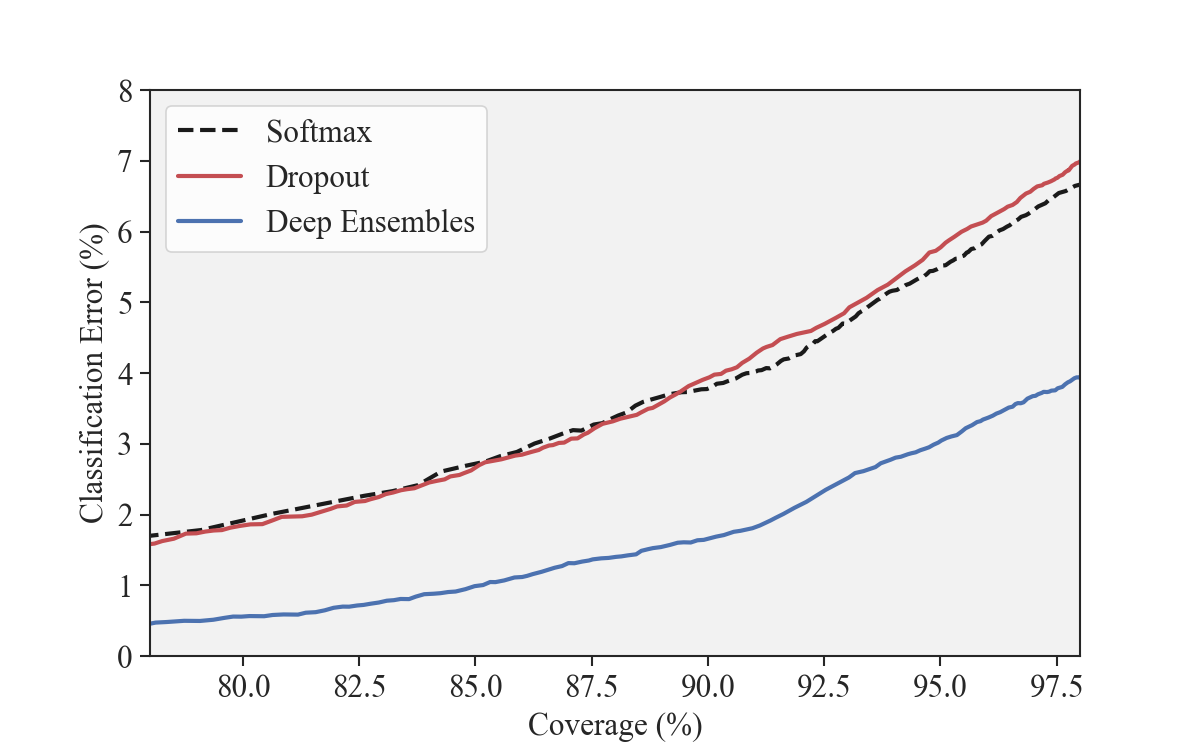}
    \caption{\textbf{Uncertainty-guided decision referral.} Risk coverage curves for different DNN approaches.}
    \label{select}
\end{figure}

Next, we tested another practical use of predictive uncertainties to identify confusing samples so that DNN can be refrained from making a prediction. This reduces the risk of making erroneous decision by rejecting to trust the classifier on certain instances and refer other difficult material samples for further testing and evaluation, instead of making an erroneous decision. This idea, formally referred as \emph{selective classification} \cite{chow1957optimum,el2010foundations}, has been introduced recently in the context of DNN classifiers \cite{geifman2017selective}. 

In this case study, we design the reject mechanism based on the predictive entropy, where the user is allowed to reject the prediction if the entropy of a DNN prediction exceeds a certain threshold. The quality of uncertainty can then be reflected by the effectiveness of the reject mechanism. To measure the performance, we adopted the \emph{risk-coverage trade-off curve} \cite{el2010foundations,geifman2017selective} of selective classification. Holistically, the \emph{coverage} refers to the ratio of data points for which the classifier is confident enough (i.e., the predictive uncertainty is lower than a given threshold) while the \emph{risk} presents the classification error among such sufficiently-confident points. The ideal goal would be to minimize the risk while maximizing the coverage.

We compared different approaches for selective classification based on the risk-coverage trade-off curves in \autoref{select}. We see that the trade-off curves based on the baseline Softmax and Dropout uncertainties are nearly identical, while Deep Ensembles performed much better on this task. For example, while offering the same level of 90\% coverage (i.e., the classifier will reject to classify on the top 10\% instances that it is mostly uncertain about),  Deep Ensembles has around 1.5\% classification error, much lower than the other two approaches (3.5\%). This further verifies the superior uncertainty qualities of Deep Ensembles, and presents us another practical benefit of well-calibrated prediction uncertainties. 

\textbf{To summarize, our results show that Deep Ensemble uncertainty guided decision referral can dramatically improve the classification accuracy on the non-referred material samples while maintaining a minimal fraction of referred (rejected) material samples.}

\subsubsection{Case Study 3: How to make DNNs recognize Out-of-Distribution examples?}

In the real-world setting, DNNs often encounter data collected in a different condition from those being used in the DNN training process. This can occur because of (a) changes in the image acquisition conditions, (b) changes in the synthesis conditions (e.g., discovery of a new material), or (c) unrelated data sample.
In such cases, it is crucial to have a detection mechanism to flag such \emph{out-of-distribution} (OOD) data points that are far away from the training data's distribution.

In this section, we test the potential use of predictive uncertainties for detecting OOD data points, with the underlying logic being that DNN models with well-calibrated uncertainties should assign higher predictive uncertainties to the OOD instances. We formulate the OOD detection problem as a binary classification problem based on the predictive entropy, and quantify the performance of the corresponding OOD classifiers. The OOD data are regarded as the positive class and the in-distribution data as the negative class, and the OOD classifiers make decisions solely based on the values of prediction entropy. We adopt the evaluation metric in ref.~\citenum{hendrycks2018deep}, and measure the classification performance using the \emph{Receiver Operating Characteristic curve} (ROC) and the \emph{Area Under the Curve} (AUC). The ROC curve plots the False Positive Rate (the probability of in-distribution data being classified as OOD) versus the True Positive Rate (the probability of OOD data being classified as OOD) for different threshold on the entropy. Therefore, the closer the ROC curve is to the upper left corner \textcolor{black}{($0.0, 1.0$)}, the better the OOD classifier is \cite{zweig1993receiver}, while a totally uninformative classifier would exhibit a diagonal ROC curve. The AUC provides a quantitative measure on the ROC curves by measuring the areas under them. Higher (closer to 1) AUC value is better, while the uninformative classifier has an AUC of 0.5.

\paragraph{Detecting Changes in Image Acquisition Conditions.}

\begin{figure}[!t]
    \centering
    \begin{subfigure}[t]{0.8\textwidth}
        \centering
        \includegraphics[height=1.2in]{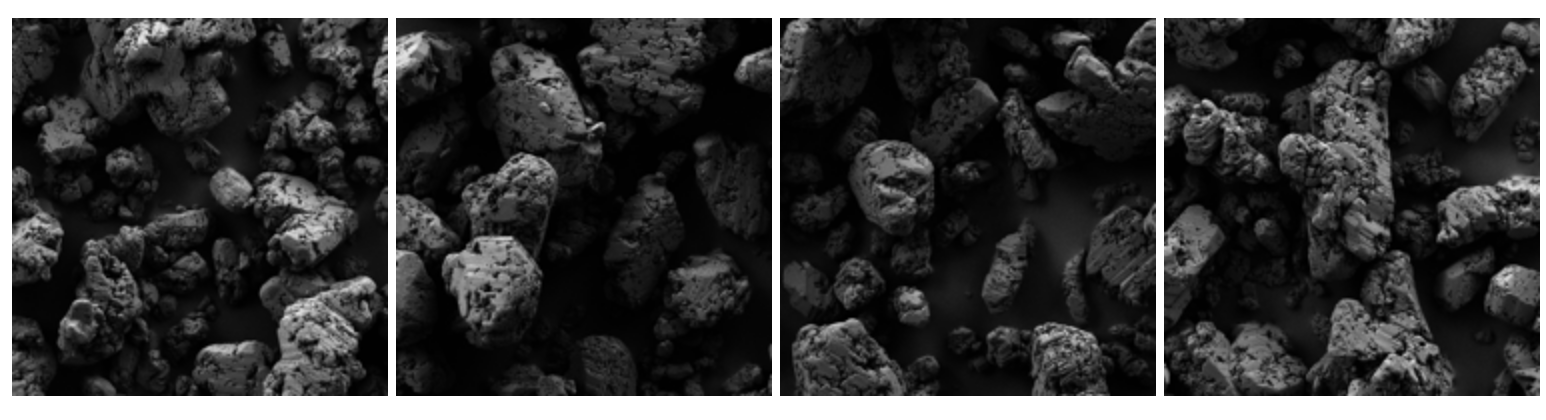}
        \caption{Representative SEM images with the original filament}
    \end{subfigure}%
    \\
    \begin{subfigure}[t]{0.8\textwidth}
        \centering
        \includegraphics[height=1.15in]{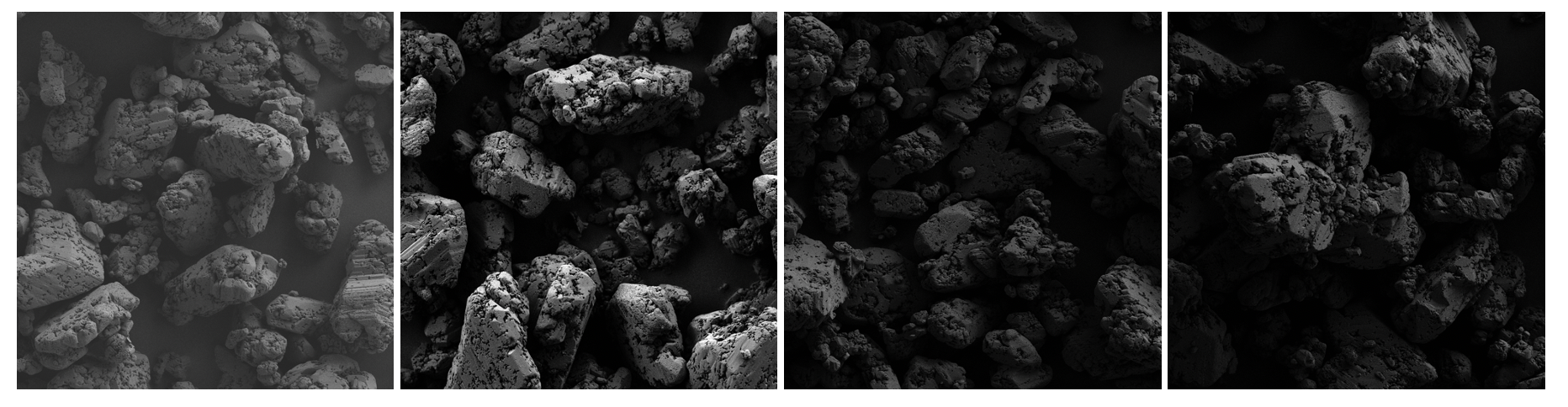}
        \caption{{\color {black} Representative SEM images with the new filament}}
    \end{subfigure}
    \caption{The effect of changing filaments on the SEM images while maintaining fixed brightness and contrast acquisition settings from the same \emph{lot}.}
    \label{change}
\end{figure}

\begin{table}[!t]
  \caption{In-distribution and OOD classification accuracy for the \emph{AT} lot}
  \label{tbl:example}
  \begin{tabular}{l|ccc}
    \hline
    Test Accuracy   & Softmax & Dropout & Deep Ensembles \\
    \hline
   In-distribution & 83.5\% & 92.1\% & 89.5\% \\
   OOD & 70.4\% &  73.1\% & 70.9\% \\
    \hline
   Drop due to OOD & 13.1\% & 19.0\% & 18.6\% \\ 
    \hline
  \end{tabular}
  \label{at_metrics}
\end{table}

To facilitate a more concrete understanding on the necessity of such OOD detection mechanism, let us first examine how our DNN classifiers perform on some real-world OOD SEM images. This particular OOD phenomena is caused by replacing the SEM filament. As a result, while the brightness and contrast settings of SEM images were held constant before and after the filament change, the newly collected images are different to the data we used in training and testing (see \autoref{change}) for the same lots. This is particularly applicable in an automated image collection workflow, where image collection conditions are usually set to static and constant conditions without human intervention. Such type of OOD is commonly denoted as \emph{covariate shift} in Machine Learning community \cite{sugiyama2007covariate}. We recorded the DNN classification accuracy on the in-distribution (original filament) and the OOD (replaced filament) SEM images from the same material lot in \autoref{at_metrics}. The gaps between in-distribution and OOD accuracy are substantial, ranging from 13\% to 19\%. This highlights the risk of being misled by DNN's erroneous predictions when encountering such real-world OOD data.

\begin{figure}[!t]
    \centering
    \begin{subfigure}[t]{0.3\textwidth}
        \centering
        \includegraphics[height=2.0in]{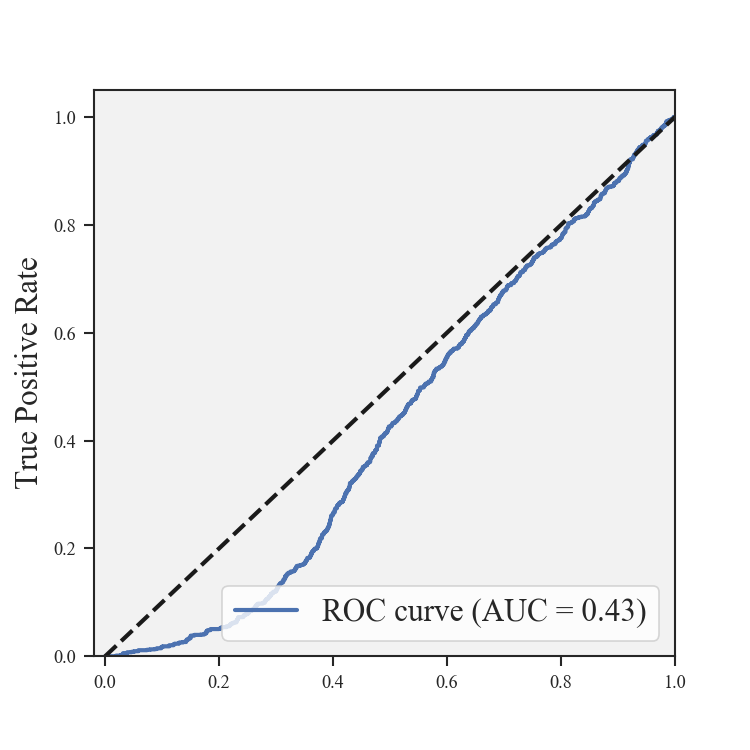}
        \caption{Softmax}
    \end{subfigure}%
    ~ 
    \begin{subfigure}[t]{0.3\textwidth}
        \centering
        \includegraphics[height=2.0in]{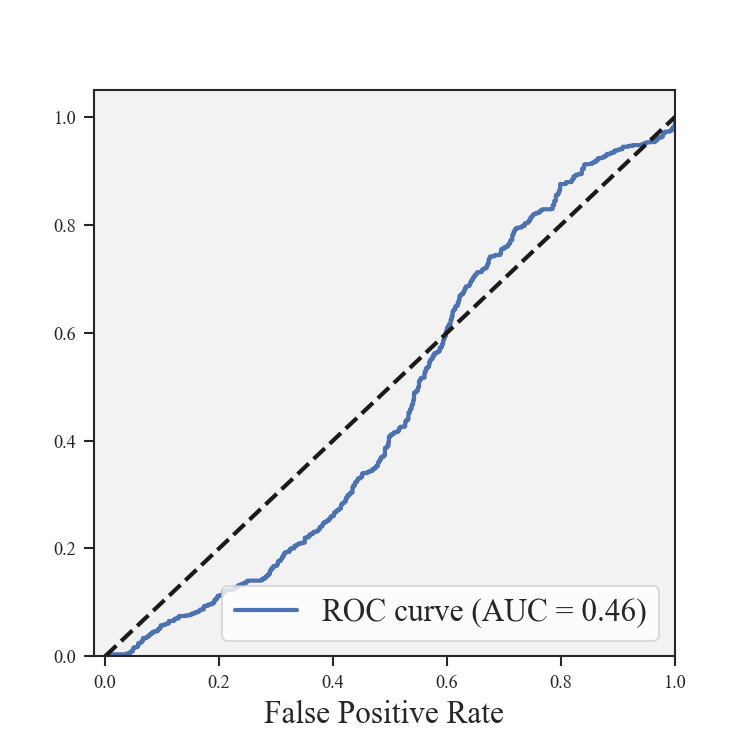}
        \caption{Dropout}
    \end{subfigure}
    ~ 
    \begin{subfigure}[t]{0.3\textwidth}
        \centering
        \includegraphics[height=2.0in]{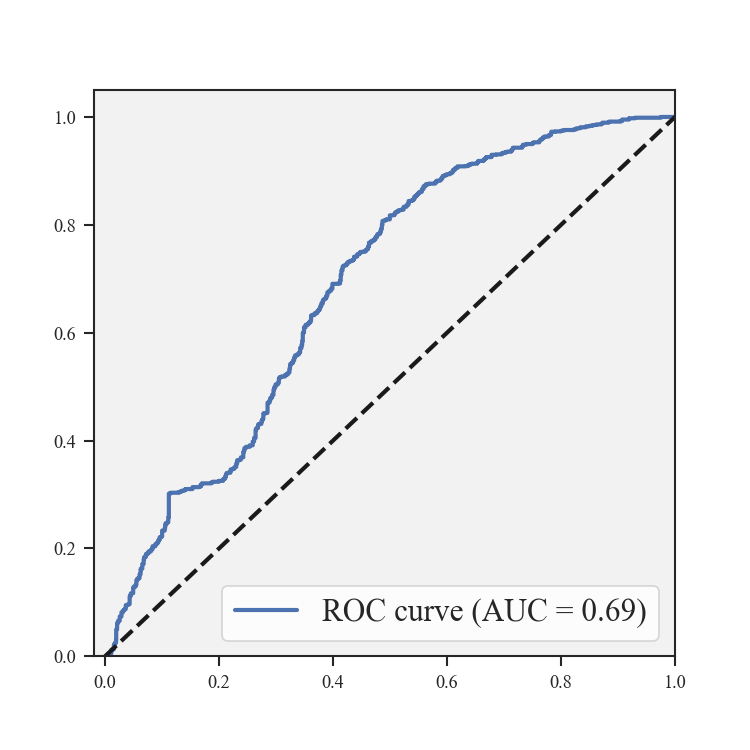}
        \caption{Deep Ensembles}
    \end{subfigure}
    \caption{ROC curves and AUC for detecting \textbf{covariate-shifted SEM material images} based on the predictive uncertainties from different approaches.}
\label{shift}
\end{figure}

Now we focus our attention to the problem of detecting the data generated using different image acquisition conditions than the training data. 
In this experiment, we use 1000 SEM images from the existing SEM image dataset as the in-distribution data, and obtain 1000 covariate shift (replaced filament) images for the same material as the OOD data. The ROC curves and AUC are shown in \autoref{shift}. The classifiers based on Softmax and Dropout uncertainties both performed poorly (flat ROC curves and low AUC value), indicating that the OOD detection based on such uncertainties will not work. This is due to the fact that the difference between in- and out-of-distribution images are subtle in this experiment, making the detection task very challenging. On the other hand, the OOD classifier based on Deep Ensembles performed much better (visually from ROC curve or quantitatively from AUC). From a practical point of view, it might be the only OOD detector capable to identify a large number of replaced-filament images without triggering a high volume of false positive alarms. The superiority of Deep Ensembles is not completely surprising -- it aligns with some prior research \cite{ovadia2019can} that also identified Deep Ensembles as the best performer on covariate shift data.

\paragraph{Detecting Changes in Synthesis Conditions.}

\begin{figure}[!t]
    \centering
    \begin{subfigure}[t]{0.3\textwidth}
        \centering
        \includegraphics[height=2.0in]{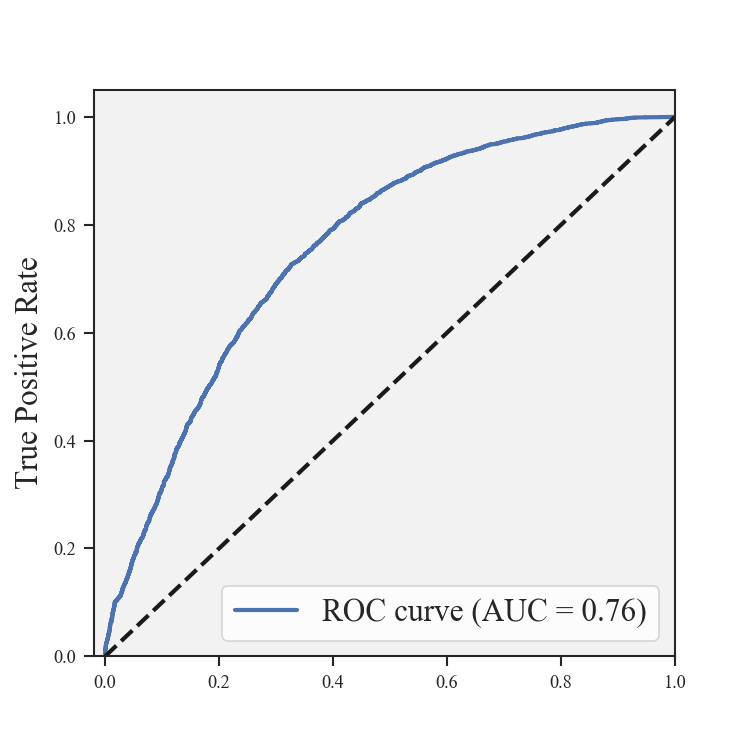}
        \caption{Softmax}
    \end{subfigure}%
    ~ 
    \begin{subfigure}[t]{0.3\textwidth}
        \centering
        \includegraphics[height=2.0in]{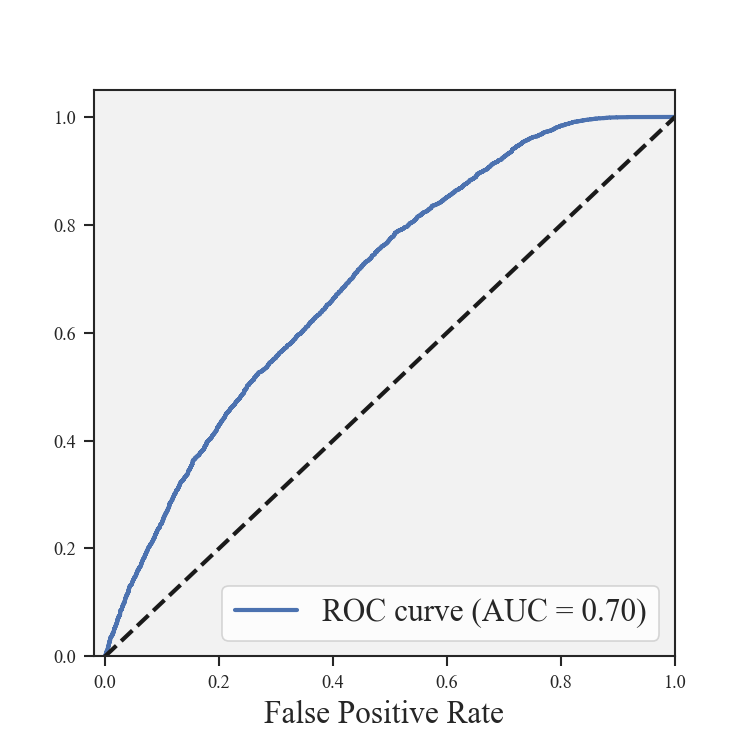}
        \caption{Dropout}
    \end{subfigure}
    ~ 
    \begin{subfigure}[t]{0.3\textwidth}
        \centering
        \includegraphics[height=2.0in]{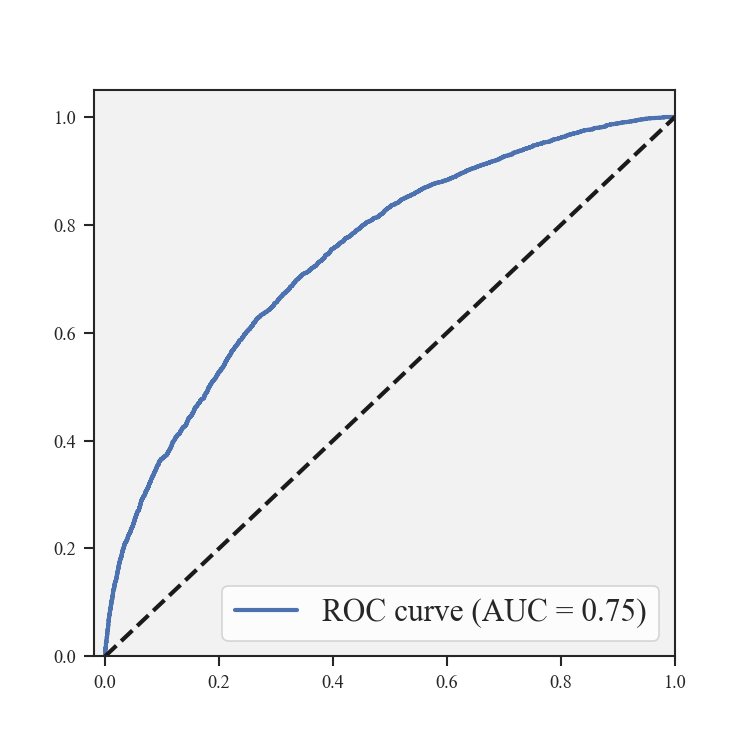}
        \caption{Deep Ensembles}
    \end{subfigure}
    \caption{ROC curves and AUC for detecting \textbf{SEM material images of unseen classes} based on the predictive uncertainties from different approaches.}
\label{novel}
\end{figure}

In this experiment, we push the OOD dataset further away from the training data distribution. Specifically, the OOD data points are SEM images for some \emph{unseen classes} of the TATB crystal material, {\color {black} i.e., they do not belong to the 30 classes in the training dataset due to different manufacturing techniques or post-processing (i.e., grinding) which will produce very different looking TATB crystals}. The occurrence of OOD data from unseen classes will be frequently encountered in realistic applications such as material discovery due to synthesis condition changes. As seen from \autoref{novel}, all examined approaches achieve acceptable performance (AUC higher than 0.7), meaning that they should be applicable to distinguish the SEM images from novel material classes.

\begin{figure}[!t]

    \centering
    \begin{subfigure}[t]{0.3\textwidth}
        \centering
        \includegraphics[height=2.0in]{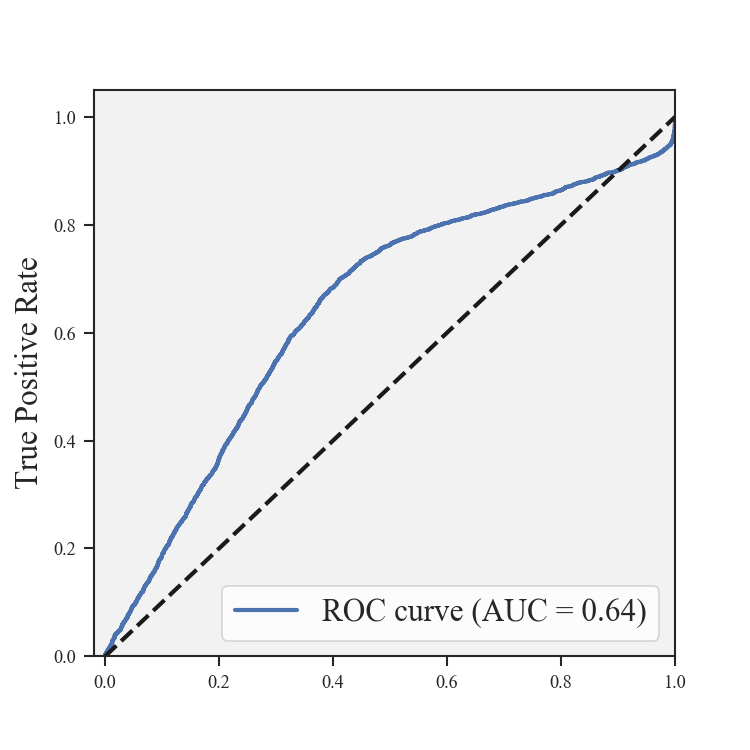}
        \caption{Softmax}
    \end{subfigure}%
    ~ 
    \begin{subfigure}[t]{0.3\textwidth}
        \centering
        \includegraphics[height=2.0in]{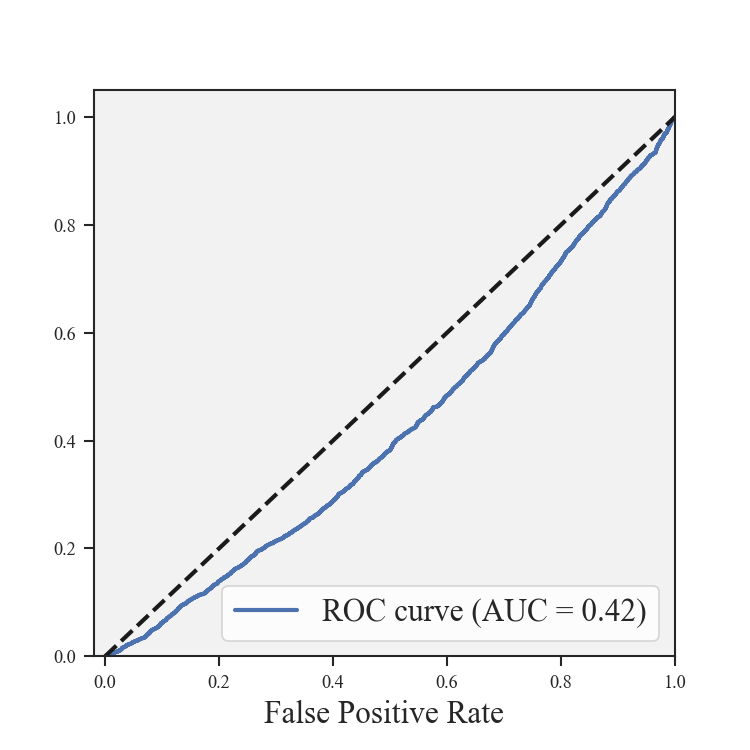}
        \caption{Dropout}
    \end{subfigure}
    ~ 
    \begin{subfigure}[t]{0.3\textwidth}
        \centering
        \includegraphics[height=2.0in]{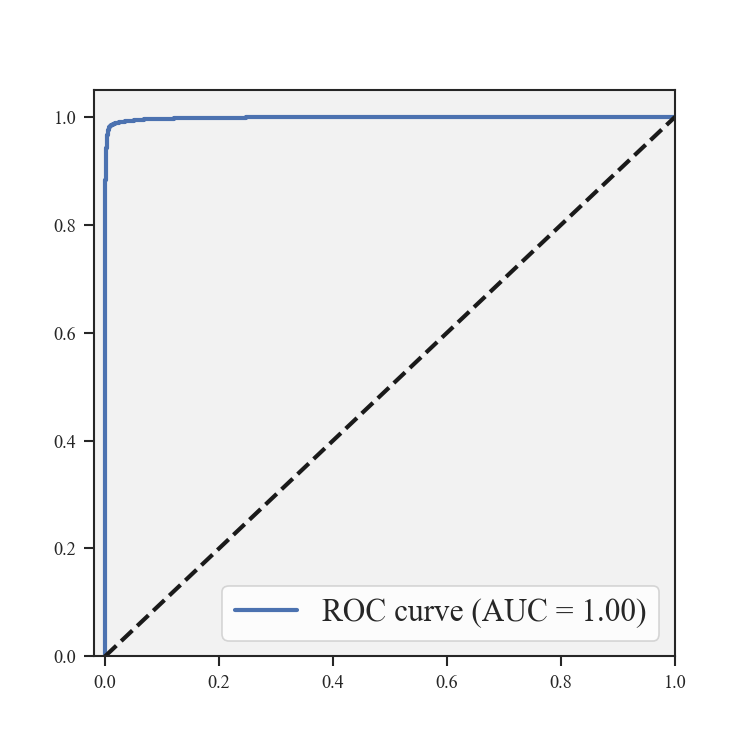}
        \caption{Deep Ensembles}
    \end{subfigure}
    \caption{ROC curves and AUC for detecting \textbf{unrelated CIFAR-10 images} based on the predictive uncertainties from different approaches.}
\label{cifar}
\end{figure}

\paragraph{Detecting Unrelated Data.}
Finally, we examine an extreme case for OOD detection, where the OOD samples are truly far away (or unrelated) from the distribution of material SEM images. For this case, we obtained OOD images from the CIFAR-10 natural image dataset (including 10 categories of images, such as cats, dogs and birds) \cite{krizhevsky2009learning}. As the CIFAR-10 images were RGB color images on lower (32 by 32) resolution, some grayscale transformation and upsampling was conducted to convert them into the format of SEM images (64 by 64 grayscale). The detection results are shown in \autoref{cifar}. Deep Ensembles achieved a near-perfect detection result (AUC close to 1). Interestingly, the simple Softmax baseline also performed better than Dropout, as the latter only achieved a 0.42 AUC (worse than randomly guessing). Although CIFAR-10 images are visually distinguishable than material SEM images, the results show that it is still non-trivial to obtain a good uncertainty-based OOD detector. The near-perfect performance of Deep Ensembles is very impressive and validates the superiority of its uncertainty estimates over Softmax and Dropout.

\begin{figure}[!t]
    \centering
    \begin{subfigure}[t]{0.3\textwidth}
        \centering
        \includegraphics[height=2.0in]{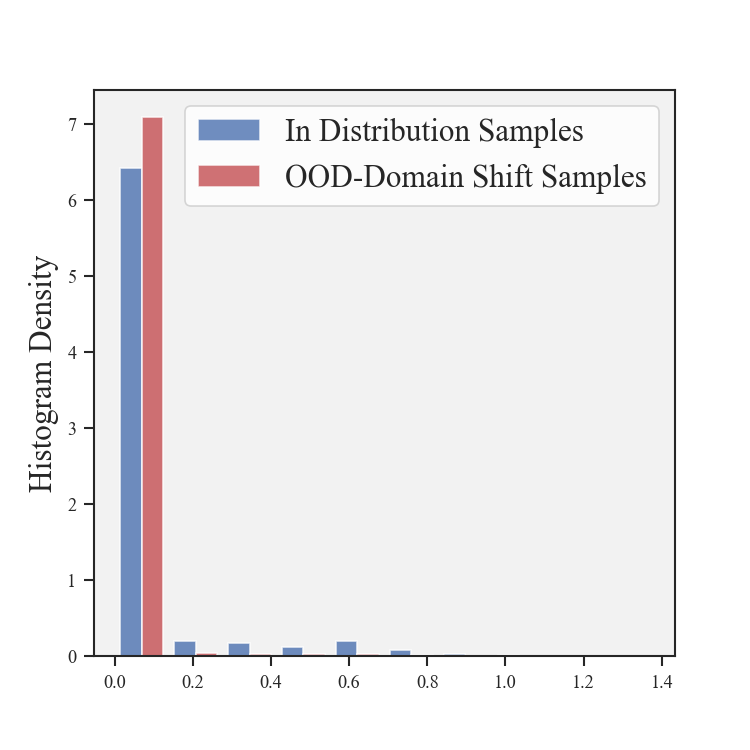}
    \end{subfigure}%
    ~ 
    \begin{subfigure}[t]{0.3\textwidth}
        \centering
        \includegraphics[height=2.0in]{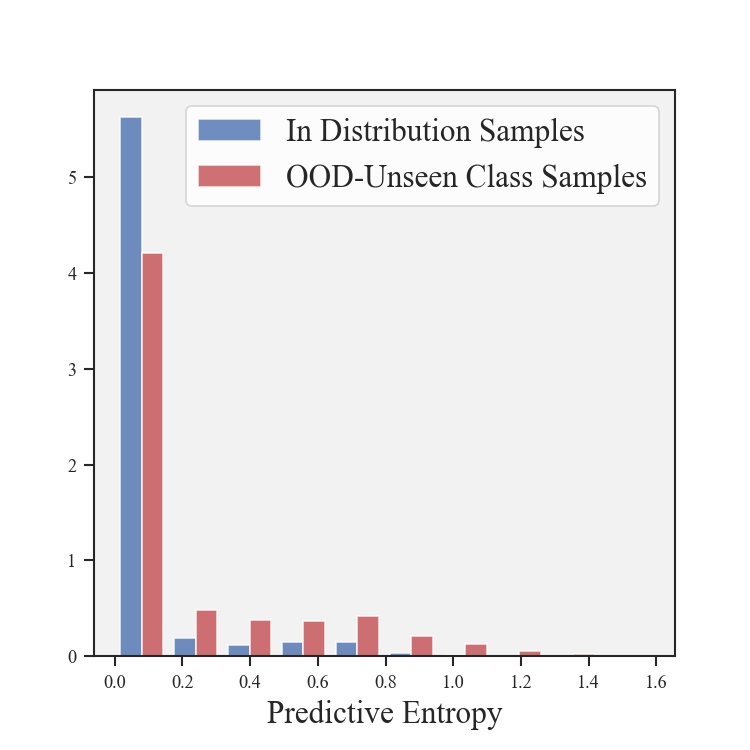}
    \end{subfigure}
    ~ 
    \begin{subfigure}[t]{0.3\textwidth}
        \centering
        \includegraphics[height=2.0in]{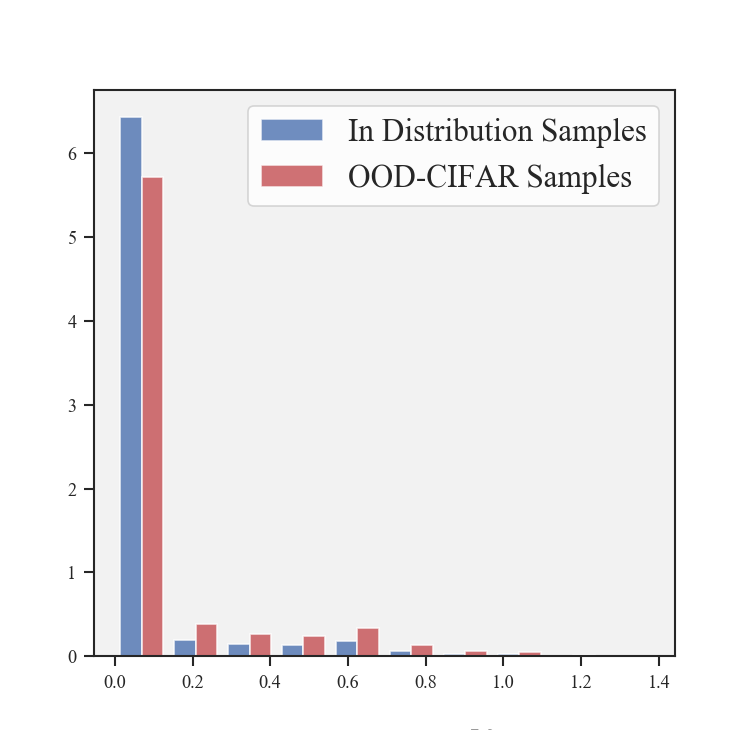}
    \end{subfigure}
    \caption{\textbf{Softmax:} histogram comparisons of the predictive entropy for the in-distribution and out-of-distributions from various datasets.}
\label{histo_at}
\end{figure}

\begin{figure}[!t]

    \centering
    \begin{subfigure}[t]{0.3\textwidth}
        \centering
        \includegraphics[height=2.0in]{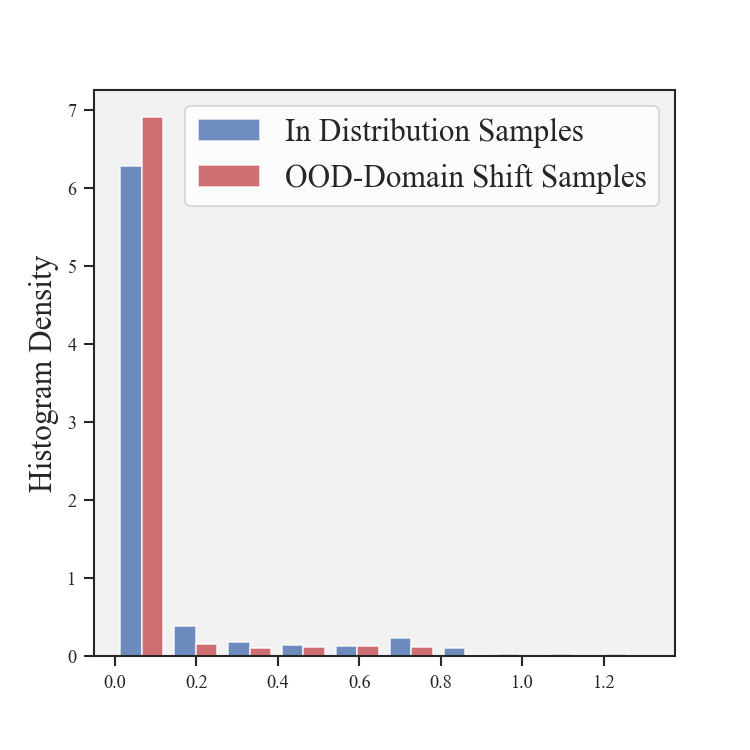}
    \end{subfigure}%
    ~ 
    \begin{subfigure}[t]{0.3\textwidth}
        \centering
        \includegraphics[height=2.0in]{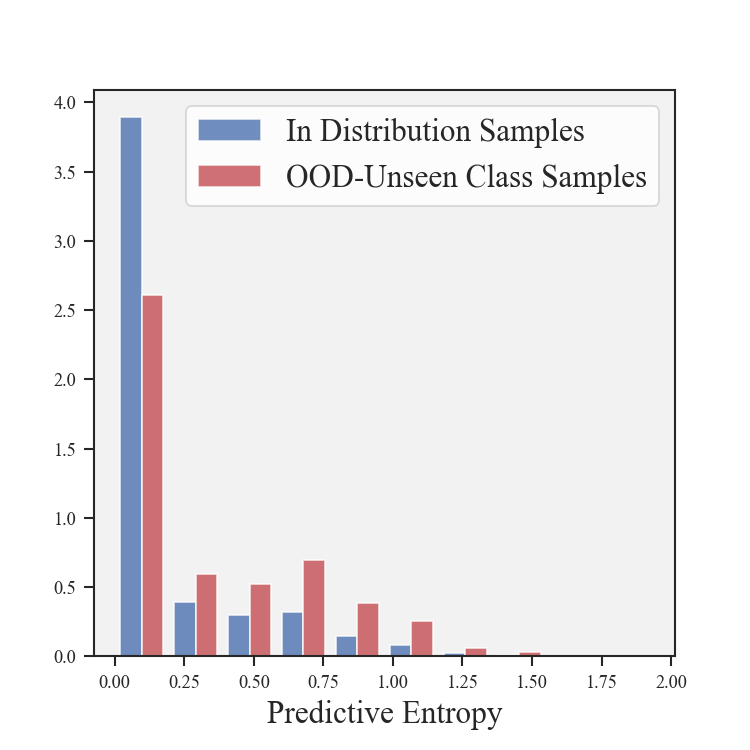}
    \end{subfigure}
    ~ 
    \begin{subfigure}[t]{0.3\textwidth}
        \centering
        \includegraphics[height=2.0in]{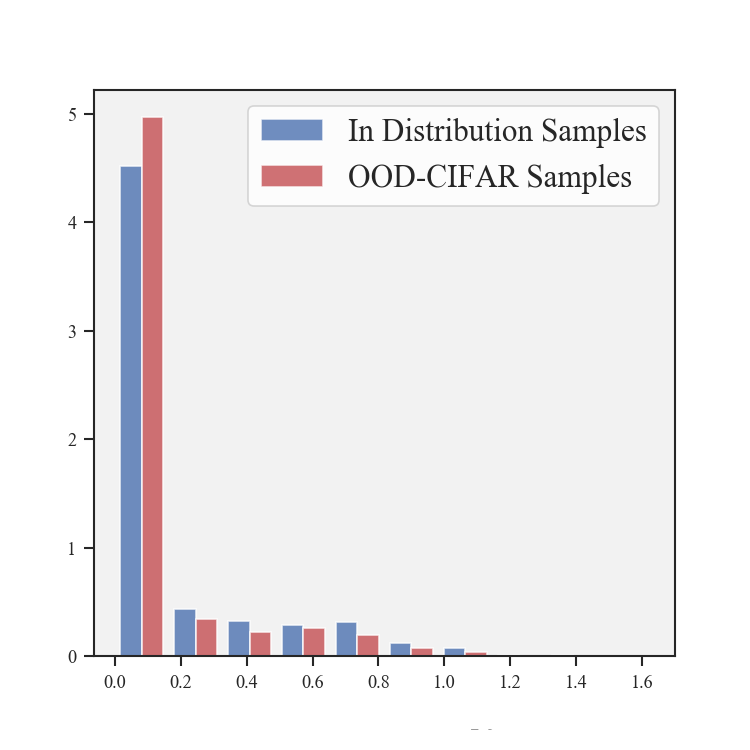}
    \end{subfigure}
    \caption{\textbf{Dropout:} histogram comparisons of the predictive entropy for the in-distribution and out-of-distributions from various datasets.}
\label{histo_dropout}
\end{figure}

\begin{figure}[!t]

    \centering
    \begin{subfigure}[t]{0.3\textwidth}
        \centering
        \includegraphics[height=2.0in]{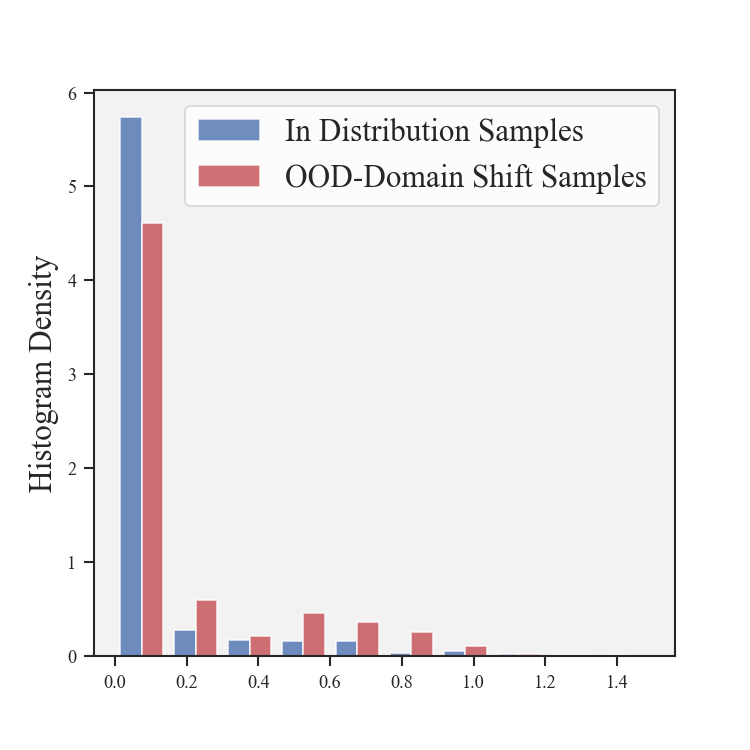}
    \end{subfigure}%
    ~ 
    \begin{subfigure}[t]{0.3\textwidth}
        \centering
        \includegraphics[height=2.0in]{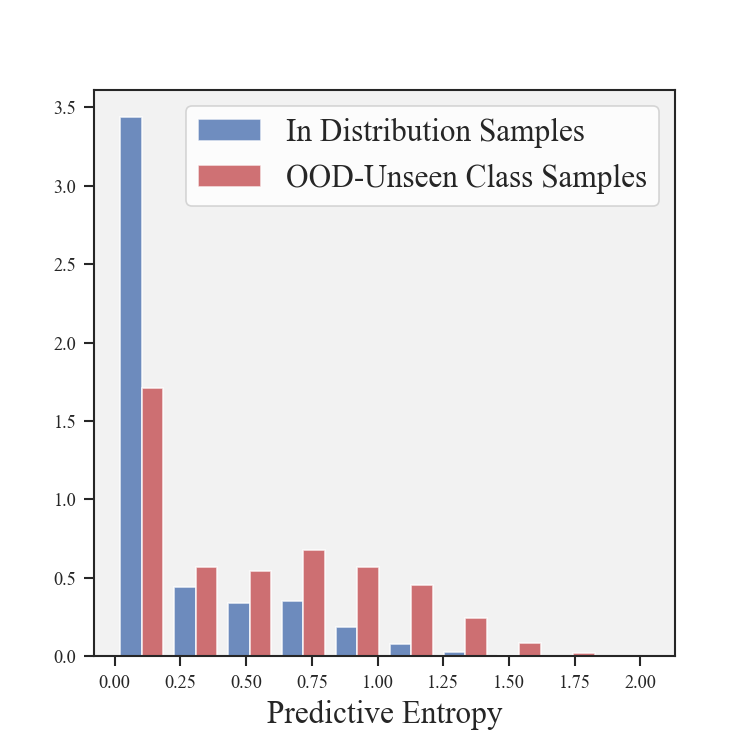}
    \end{subfigure}
    ~ 
    \begin{subfigure}[t]{0.3\textwidth}
        \centering
        \includegraphics[height=2.0in]{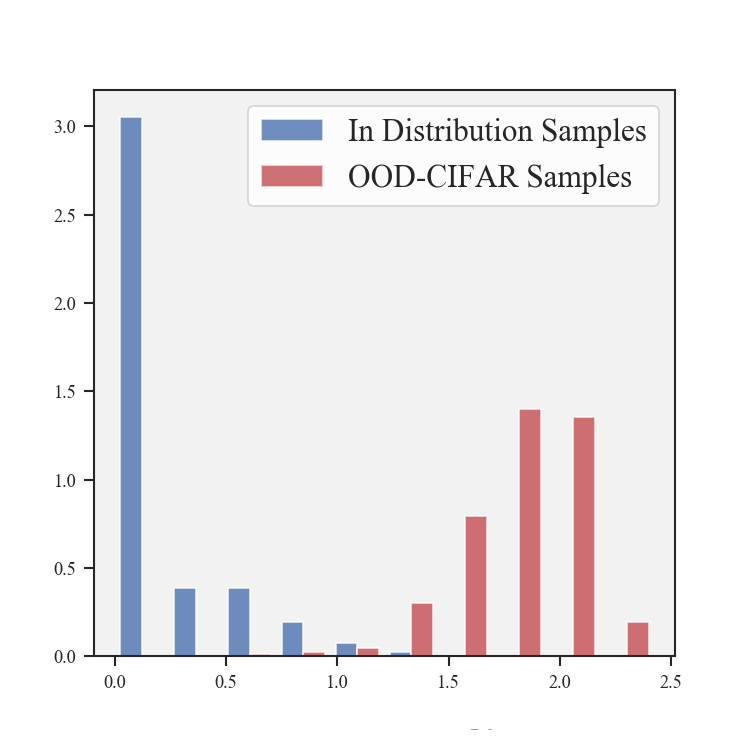}
    \end{subfigure}
    \caption{\textbf{Deep Ensembles:} histogram comparisons of the predictive entropy for the in-distribution and out-of-distributions from various datasets.}
\label{histo_de}
\end{figure}
\paragraph{Can we leverage uncertainties to identify different types of shifts?}
In this section, we ask the question if uncertainty-guided OOD detection approaches can differentiate among different sources of distributions shifts. This is an important feature to have as this might inform users on what should they do with the OOD data, for example, if the user could utilize the OOD data to augment the existing training data (the case of changing image acquisition conditions), conduct further testing (the case of changing synthesis conditions), or simply discard the data (the case of unrelated data).  

To answer this question, we characterize the distribution of predictive entropy for in-distribution and OOD data using histograms in \autoref{histo_at} to \autoref{histo_de}. Intuitively, we expect the predictive entropy of OOD data to be always higher (i.e., more uncertain) than in-distribution ones. Furthermore, this discrepancy should become more noticeable as the OOD data shifts away from the training data distribution. However, from \autoref{histo_at} and \autoref{histo_dropout}, we observe that both Softmax and Dropout are very confident (assigning low entropy values) on their predictions for domain-shift and CIFAR OOD data, although they both performed reasonably well for Unseen-class data. 
On the other hand, as seen in \autoref{histo_de}, Deep Ensembles always produce higher predictive entropy for all examined OOD datasets, and the gap between in-distribution and OOD samples' predictive entropy indeed becomes more apparent with an increase in the amount of shift. In other words, Deep Ensembles can differentiate among different sources of shifts.

\textbf{To summarize, our results show that uncertainties from Deep Ensembles can be used to detect out-of-distribution sample. Further, their uncertainties are able to differentiate the sources of distribution shifts and hint towards what to do with the OOD data, e.g., using the OOD data with changed image acquisition conditions in data augmentation, conducting new mechanical testing after detecting the OOD data from unseen classes of materials, or simply discarding the unrelated OOD data.}

\section{Conclusion}
\label{sec:discussion}
In this work, we successfully demonstrated the benefits, applicability and limitations of uncertainty-aware deep learning methods for making materials discovery workflows more dependable. Specifically, we showed how uncertainty-guided methods can serve as a unified approach to answer several important issues in {\color {black} the examined material classification problem}. There are still some issues yet to be resolved for a successful application of machine learning in Materials Discovery workflows, but leveraging uncertainties in DL models is a first step to addressing implementation of DL models for materials applications.

\begin{acknowledgement}
This work was performed under the auspices of the U.S. Department of Energy by Lawrence Livermore National Laboratory under Contract DE-AC52-07NA27344 and was supported by the LLNL-LDRD Program under Project No. 19-SI-001. This work is reviewed and released under LLNL-JRNL-816936. The authors thank Brian Gallagher for providing valuable discussions and information.
\end{acknowledgement}





\section{Author Information}

\subsection{Corresponding Authors}
\begin{itemize}
    \item Jize Zhang, Center for Applied Scientific Computing, Computing Directorate, Lawrence Livermore National Laboratory, Livermore, California 94550, United States, Email: zhang64@llnl.gov
    \item T. Yong-Jin Han, Materials Science Division, Physical and Life Sciences Directorate, Lawrence Livermore National Laboratory, Livermore, California 94550, United States, Email: han5@llnl.gov
\end{itemize}

\subsection{Authors}
\begin{itemize}
\item Bhavya Kailkhura, Center for Applied Scientific Computing, Computing Directorate, Lawrence Livermore National Laboratory, Livermore, California 94550, United States; Email: kailkhura1@llnl.gov
\end{itemize}

\bibliography{acs-short}

\providecommand{\latin}[1]{#1}
\makeatletter
\providecommand{\doi}
  {\begingroup\let\do\@makeother\dospecials
  \catcode`\{=1 \catcode`\}=2 \doi@aux}
\providecommand{\doi@aux}[1]{\endgroup\texttt{#1}}
\makeatother
\providecommand*\mcitethebibliography{\thebibliography}
\csname @ifundefined\endcsname{endmcitethebibliography}
  {\let\endmcitethebibliography\endthebibliography}{}
\begin{mcitethebibliography}{48}
\providecommand*\natexlab[1]{#1}
\providecommand*\mciteSetBstSublistMode[1]{}
\providecommand*\mciteSetBstMaxWidthForm[2]{}
\providecommand*\mciteBstWouldAddEndPuncttrue
  {\def\EndOfBibitem{\unskip.}}
\providecommand*\mciteBstWouldAddEndPunctfalse
  {\let\EndOfBibitem\relax}
\providecommand*\mciteSetBstMidEndSepPunct[3]{}
\providecommand*\mciteSetBstSublistLabelBeginEnd[3]{}
\providecommand*\EndOfBibitem{}
\mciteSetBstSublistMode{f}
\mciteSetBstMaxWidthForm{subitem}{(\alph{mcitesubitemcount})}
\mciteSetBstSublistLabelBeginEnd
  {\mcitemaxwidthsubitemform\space}
  {\relax}
  {\relax}

\bibitem[Raghu and Schmidt(2020)Raghu, and Schmidt]{raghu2020survey}
Raghu,~M.; Schmidt,~E. A Survey of Deep Learning for Scientific Discovery.
  \textbf{2020}, arXiv: 2003.11755 [cs.LG]\relax
\mciteBstWouldAddEndPuncttrue
\mciteSetBstMidEndSepPunct{\mcitedefaultmidpunct}
{\mcitedefaultendpunct}{\mcitedefaultseppunct}\relax
\EndOfBibitem
\bibitem[Ramprasad \latin{et~al.}(2017)Ramprasad, Batra, Pilania,
  Mannodi-Kanakkithodi, and Kim]{ramprasad2017machine}
Ramprasad,~R.; Batra,~R.; Pilania,~G.; Mannodi-Kanakkithodi,~A.; Kim,~C.
  Machine learning in materials informatics: recent applications and prospects.
  \emph{npj Comput. Mater.} \textbf{2017}, \emph{3}, 1--13\relax
\mciteBstWouldAddEndPuncttrue
\mciteSetBstMidEndSepPunct{\mcitedefaultmidpunct}
{\mcitedefaultendpunct}{\mcitedefaultseppunct}\relax
\EndOfBibitem
\bibitem[Kailkhura \latin{et~al.}(2019)Kailkhura, Gallagher, Kim, Hiszpanski,
  and Han]{kailkhura2019reliable}
Kailkhura,~B.; Gallagher,~B.; Kim,~S.; Hiszpanski,~A.; Han,~T. Y.-J. Reliable
  and explainable machine-learning methods for accelerated material discovery.
  \emph{npj Comput. Mater.} \textbf{2019}, \emph{5}, 1--9\relax
\mciteBstWouldAddEndPuncttrue
\mciteSetBstMidEndSepPunct{\mcitedefaultmidpunct}
{\mcitedefaultendpunct}{\mcitedefaultseppunct}\relax
\EndOfBibitem
\bibitem[Gallagher \latin{et~al.}(2020)Gallagher, Rever, Loveland, Mundhenk,
  Beauchamp, Robertson, Jaman, Hiszpanski, and Han]{gallagher2020predicting}
Gallagher,~B.; Rever,~M.; Loveland,~D.; Mundhenk,~T.~N.; Beauchamp,~B.;
  Robertson,~E.; Jaman,~G.~G.; Hiszpanski,~A.~M.; Han,~T. Y.-J. Predicting
  compressive strength of consolidated molecular solids using computer vision
  and deep learning. \emph{Mater. Des.} \textbf{2020}, 108541\relax
\mciteBstWouldAddEndPuncttrue
\mciteSetBstMidEndSepPunct{\mcitedefaultmidpunct}
{\mcitedefaultendpunct}{\mcitedefaultseppunct}\relax
\EndOfBibitem
\bibitem[Ling \latin{et~al.}(2017)Ling, Hutchinson, Antono, DeCost, Holm, and
  Meredig]{ling2017building}
Ling,~J.; Hutchinson,~M.; Antono,~E.; DeCost,~B.; Holm,~E.~A.; Meredig,~B.
  Building data-driven models with microstructural images: Generalization and
  interpretability. \emph{Materials Discovery} \textbf{2017}, \emph{10},
  19--28\relax
\mciteBstWouldAddEndPuncttrue
\mciteSetBstMidEndSepPunct{\mcitedefaultmidpunct}
{\mcitedefaultendpunct}{\mcitedefaultseppunct}\relax
\EndOfBibitem
\bibitem[Liu \latin{et~al.}(2017)Liu, Zhao, Ju, and Shi]{liu2017materials}
Liu,~Y.; Zhao,~T.; Ju,~W.; Shi,~S. Materials discovery and design using machine
  learning. \emph{Journal of Materiomics} \textbf{2017}, \emph{3},
  159--177\relax
\mciteBstWouldAddEndPuncttrue
\mciteSetBstMidEndSepPunct{\mcitedefaultmidpunct}
{\mcitedefaultendpunct}{\mcitedefaultseppunct}\relax
\EndOfBibitem
\bibitem[Haghighatlari \latin{et~al.}(2019)Haghighatlari, Shih, and
  Hachmann]{haghighatlari2019thinking}
Haghighatlari,~M.; Shih,~C.-Y.; Hachmann,~J. Thinking globally, acting locally:
  on the issue of training set imbalance and the case for local machine
  learning models in chemistry. \textbf{2019}, chemrXiv: 8796947\relax
\mciteBstWouldAddEndPuncttrue
\mciteSetBstMidEndSepPunct{\mcitedefaultmidpunct}
{\mcitedefaultendpunct}{\mcitedefaultseppunct}\relax
\EndOfBibitem
\bibitem[{Bulusu} \latin{et~al.}(2020){Bulusu}, {Kailkhura}, {Li}, {Varshney},
  and {Song}]{Bulusu20}
{Bulusu},~S.; {Kailkhura},~B.; {Li},~B.; {Varshney},~P.~K.; {Song},~D.
  Anomalous Example Detection in Deep Learning: A Survey. \emph{IEEE Access}
  \textbf{2020}, \emph{8}, 132330--132347\relax
\mciteBstWouldAddEndPuncttrue
\mciteSetBstMidEndSepPunct{\mcitedefaultmidpunct}
{\mcitedefaultendpunct}{\mcitedefaultseppunct}\relax
\EndOfBibitem
\bibitem[Cho \latin{et~al.}(2015)Cho, Lee, Shin, Choy, and Do]{cho2015much}
Cho,~J.; Lee,~K.; Shin,~E.; Choy,~G.; Do,~S. How much data is needed to train a
  medical image deep learning system to achieve necessary high accuracy?
  \textbf{2015}, arXiv: 1511.06348 [cs.LG]\relax
\mciteBstWouldAddEndPuncttrue
\mciteSetBstMidEndSepPunct{\mcitedefaultmidpunct}
{\mcitedefaultendpunct}{\mcitedefaultseppunct}\relax
\EndOfBibitem
\bibitem[Niculescu-Mizil and Caruana(2005)Niculescu-Mizil, and
  Caruana]{niculescu2005predicting}
Niculescu-Mizil,~A.; Caruana,~R. Predicting good probabilities with supervised
  learning. International Conference on Machine Learning. 2005; pp
  625--632\relax
\mciteBstWouldAddEndPuncttrue
\mciteSetBstMidEndSepPunct{\mcitedefaultmidpunct}
{\mcitedefaultendpunct}{\mcitedefaultseppunct}\relax
\EndOfBibitem
\bibitem[Guo \latin{et~al.}(2017)Guo, Pleiss, Sun, and
  Weinberger]{guo2017calibration}
Guo,~C.; Pleiss,~G.; Sun,~Y.; Weinberger,~K.~Q. On calibration of modern neural
  networks. International Conference on Machine Learning. 2017; pp
  1321--1330\relax
\mciteBstWouldAddEndPuncttrue
\mciteSetBstMidEndSepPunct{\mcitedefaultmidpunct}
{\mcitedefaultendpunct}{\mcitedefaultseppunct}\relax
\EndOfBibitem
\bibitem[Zhang \latin{et~al.}(2020)Zhang, Kailkhura, and Han]{zhang2020mix}
Zhang,~J.; Kailkhura,~B.; Han,~T. Mix-n-Match: Ensemble and Compositional
  Methods for Uncertainty Calibration in Deep Learning. International
  Conference on Machine Learning. 2020; pp 11117--11128\relax
\mciteBstWouldAddEndPuncttrue
\mciteSetBstMidEndSepPunct{\mcitedefaultmidpunct}
{\mcitedefaultendpunct}{\mcitedefaultseppunct}\relax
\EndOfBibitem
\bibitem[Fort \latin{et~al.}(2019)Fort, Hu, and Lakshminarayanan]{fort2019deep}
Fort,~S.; Hu,~H.; Lakshminarayanan,~B. Deep ensembles: A loss landscape
  perspective. \textbf{2019}, arXiv: 1912.02757 [cs.LG]\relax
\mciteBstWouldAddEndPuncttrue
\mciteSetBstMidEndSepPunct{\mcitedefaultmidpunct}
{\mcitedefaultendpunct}{\mcitedefaultseppunct}\relax
\EndOfBibitem
\bibitem[Ovadia \latin{et~al.}(2019)Ovadia, Fertig, Ren, Nado, Sculley,
  Nowozin, Dillon, Lakshminarayanan, and Snoek]{ovadia2019can}
Ovadia,~Y.; Fertig,~E.; Ren,~J.; Nado,~Z.; Sculley,~D.; Nowozin,~S.;
  Dillon,~J.; Lakshminarayanan,~B.; Snoek,~J. Can you trust your model's
  uncertainty? Evaluating predictive uncertainty under dataset shift. Advances
  in Neural Information Processing Systems. 2019; pp 13991--14002\relax
\mciteBstWouldAddEndPuncttrue
\mciteSetBstMidEndSepPunct{\mcitedefaultmidpunct}
{\mcitedefaultendpunct}{\mcitedefaultseppunct}\relax
\EndOfBibitem
\bibitem[Hein \latin{et~al.}(2019)Hein, Andriushchenko, and
  Bitterwolf]{hein2019relu}
Hein,~M.; Andriushchenko,~M.; Bitterwolf,~J. Why relu networks yield
  high-confidence predictions far away from the training data and how to
  mitigate the problem. Proceedings of the IEEE/CVF Conference on Computer
  Vision and Pattern Recognition. 2019; pp 41--50\relax
\mciteBstWouldAddEndPuncttrue
\mciteSetBstMidEndSepPunct{\mcitedefaultmidpunct}
{\mcitedefaultendpunct}{\mcitedefaultseppunct}\relax
\EndOfBibitem
\bibitem[Mallick \latin{et~al.}(2020)Mallick, Dwivedi, Kailkhura, Joshi, and
  Han]{mallick2020probabilistic}
Mallick,~A.; Dwivedi,~C.; Kailkhura,~B.; Joshi,~G.; Han,~T. Probabilistic
  Neighbourhood Component Analysis: Sample Efficient Uncertainty Estimation in
  Deep Learning. \textbf{2020}, arXiv: 2007.10800 [cs.LG]\relax
\mciteBstWouldAddEndPuncttrue
\mciteSetBstMidEndSepPunct{\mcitedefaultmidpunct}
{\mcitedefaultendpunct}{\mcitedefaultseppunct}\relax
\EndOfBibitem
\bibitem[Willey \latin{et~al.}(2006)Willey, van Buuren, Lee, Overturf, Kinney,
  Handly, Weeks, and Ilavsky]{willey2006changes}
Willey,~T.~M.; van Buuren,~T.; Lee,~J.~R.; Overturf,~G.~E.; Kinney,~J.~H.;
  Handly,~J.; Weeks,~B.~L.; Ilavsky,~J. Changes in Pore Size Distribution upon
  Thermal Cycling of TATB-based Explosives Measured by Ultra-Small Angle X-Ray
  Scattering. \emph{Propellants, Explos., Pyrotech.} \textbf{2006}, \emph{31},
  466--471\relax
\mciteBstWouldAddEndPuncttrue
\mciteSetBstMidEndSepPunct{\mcitedefaultmidpunct}
{\mcitedefaultendpunct}{\mcitedefaultseppunct}\relax
\EndOfBibitem
\bibitem[Gal and Ghahramani(2016)Gal, and Ghahramani]{gal2016dropout}
Gal,~Y.; Ghahramani,~Z. Dropout as a Bayesian approximation: Representing model
  uncertainty in deep learning. International Conference on Machine Learning.
  2016; pp 1651--1660\relax
\mciteBstWouldAddEndPuncttrue
\mciteSetBstMidEndSepPunct{\mcitedefaultmidpunct}
{\mcitedefaultendpunct}{\mcitedefaultseppunct}\relax
\EndOfBibitem
\bibitem[Louizos and Welling(2017)Louizos, and
  Welling]{louizos2017multiplicative}
Louizos,~C.; Welling,~M. Multiplicative Normalizing Flows for Variational
  Bayesian Neural Networks. International Conference on Machine Learning. 2017;
  pp 2218--2227\relax
\mciteBstWouldAddEndPuncttrue
\mciteSetBstMidEndSepPunct{\mcitedefaultmidpunct}
{\mcitedefaultendpunct}{\mcitedefaultseppunct}\relax
\EndOfBibitem
\bibitem[Kuleshov \latin{et~al.}(2018)Kuleshov, Fenner, and
  Ermon]{kuleshov2018accurate}
Kuleshov,~V.; Fenner,~N.; Ermon,~S. Accurate Uncertainties for Deep Learning
  Using Calibrated Regression. International Conference on Machine Learning.
  2018; pp 2796--2804\relax
\mciteBstWouldAddEndPuncttrue
\mciteSetBstMidEndSepPunct{\mcitedefaultmidpunct}
{\mcitedefaultendpunct}{\mcitedefaultseppunct}\relax
\EndOfBibitem
\bibitem[Cort{\'e}s-Ciriano and Bender(2019)Cort{\'e}s-Ciriano, and
  Bender]{cortes2019reliable}
Cort{\'e}s-Ciriano,~I.; Bender,~A. Reliable prediction errors for deep neural
  networks using test-time dropout. \emph{J. Chem. Inf. Model.} \textbf{2019},
  \emph{59}, 3330--3339\relax
\mciteBstWouldAddEndPuncttrue
\mciteSetBstMidEndSepPunct{\mcitedefaultmidpunct}
{\mcitedefaultendpunct}{\mcitedefaultseppunct}\relax
\EndOfBibitem
\bibitem[Lakshminarayanan \latin{et~al.}(2017)Lakshminarayanan, Pritzel, and
  Blundell]{lakshminarayanan2017simple}
Lakshminarayanan,~B.; Pritzel,~A.; Blundell,~C. Simple and scalable predictive
  uncertainty estimation using deep ensembles. Advances in Neural Information
  Processing Systems. 2017; pp 6402--6413\relax
\mciteBstWouldAddEndPuncttrue
\mciteSetBstMidEndSepPunct{\mcitedefaultmidpunct}
{\mcitedefaultendpunct}{\mcitedefaultseppunct}\relax
\EndOfBibitem
\bibitem[Zagoruyko and Komodakis(2016)Zagoruyko, and
  Komodakis]{zagoruyko2016wide}
Zagoruyko,~S.; Komodakis,~N. Wide residual networks. Proceedings of the British
  Machine Vision Conference. 2016; pp 87.1--87.12\relax
\mciteBstWouldAddEndPuncttrue
\mciteSetBstMidEndSepPunct{\mcitedefaultmidpunct}
{\mcitedefaultendpunct}{\mcitedefaultseppunct}\relax
\EndOfBibitem
\bibitem[Rawat and Wang(2017)Rawat, and Wang]{rawat2017deep}
Rawat,~W.; Wang,~Z. Deep convolutional neural networks for image
  classification: A comprehensive review. \emph{Neural Computation}
  \textbf{2017}, \emph{29}, 2352--2449\relax
\mciteBstWouldAddEndPuncttrue
\mciteSetBstMidEndSepPunct{\mcitedefaultmidpunct}
{\mcitedefaultendpunct}{\mcitedefaultseppunct}\relax
\EndOfBibitem
\bibitem[Kingma and Ba(2015)Kingma, and Ba]{kingma2015adam}
Kingma,~D.; Ba,~J. Adam: A Method for Stochastic Optimization. International
  Conference on Learning Representations. 2015\relax
\mciteBstWouldAddEndPuncttrue
\mciteSetBstMidEndSepPunct{\mcitedefaultmidpunct}
{\mcitedefaultendpunct}{\mcitedefaultseppunct}\relax
\EndOfBibitem
\bibitem[Falkner \latin{et~al.}(2018)Falkner, Klein, and
  Hutter]{falkner2018bohb}
Falkner,~S.; Klein,~A.; Hutter,~F. BOHB: Robust and efficient hyperparameter
  optimization at scale. International Conference on Machine Learning. 2018; pp
  1437--1446\relax
\mciteBstWouldAddEndPuncttrue
\mciteSetBstMidEndSepPunct{\mcitedefaultmidpunct}
{\mcitedefaultendpunct}{\mcitedefaultseppunct}\relax
\EndOfBibitem
\bibitem[Gal(2016)]{gal2015thesis}
Gal,~Y. Uncertainty in Deep Learning. Ph.D.\ thesis, 2016\relax
\mciteBstWouldAddEndPuncttrue
\mciteSetBstMidEndSepPunct{\mcitedefaultmidpunct}
{\mcitedefaultendpunct}{\mcitedefaultseppunct}\relax
\EndOfBibitem
\bibitem[Shannon(1948)]{shannon1948mathematical}
Shannon,~C.~E. A mathematical theory of communication. \emph{The Bell System
  Technical Journal} \textbf{1948}, \emph{27}, 379--423\relax
\mciteBstWouldAddEndPuncttrue
\mciteSetBstMidEndSepPunct{\mcitedefaultmidpunct}
{\mcitedefaultendpunct}{\mcitedefaultseppunct}\relax
\EndOfBibitem
\bibitem[Gneiting and Raftery(2007)Gneiting, and Raftery]{gneiting2007strictly}
Gneiting,~T.; Raftery,~A.~E. Strictly proper scoring rules, prediction, and
  estimation. \emph{J. Am. Stat. Assoc.} \textbf{2007}, \emph{102},
  359--378\relax
\mciteBstWouldAddEndPuncttrue
\mciteSetBstMidEndSepPunct{\mcitedefaultmidpunct}
{\mcitedefaultendpunct}{\mcitedefaultseppunct}\relax
\EndOfBibitem
\bibitem[Naeini \latin{et~al.}(2015)Naeini, Cooper, and
  Hauskrecht]{naeini2015obtaining}
Naeini,~M.~P.; Cooper,~G.; Hauskrecht,~M. Obtaining well calibrated
  probabilities using bayesian binning. AAAI Conference on Artificial
  Intelligence. 2015\relax
\mciteBstWouldAddEndPuncttrue
\mciteSetBstMidEndSepPunct{\mcitedefaultmidpunct}
{\mcitedefaultendpunct}{\mcitedefaultseppunct}\relax
\EndOfBibitem
\bibitem[Dietterich(2000)]{dietterich2000ensemble}
Dietterich,~T.~G. Ensemble Methods in Machine Learning. \emph{Multiple
  Classifier Systems} \textbf{2000}, 1--15\relax
\mciteBstWouldAddEndPuncttrue
\mciteSetBstMidEndSepPunct{\mcitedefaultmidpunct}
{\mcitedefaultendpunct}{\mcitedefaultseppunct}\relax
\EndOfBibitem
\bibitem[Shi \latin{et~al.}(2018)Shi, Zhang, Liu, Cao, Ye, Cheng, and
  Zheng]{shi2018crowd}
Shi,~Z.; Zhang,~L.; Liu,~Y.; Cao,~X.; Ye,~Y.; Cheng,~M.-M.; Zheng,~G. Crowd
  counting with deep negative correlation learning. Proceedings of the IEEE
  Conference on Computer Vision and Pattern Recognition. 2018; pp
  5382--5390\relax
\mciteBstWouldAddEndPuncttrue
\mciteSetBstMidEndSepPunct{\mcitedefaultmidpunct}
{\mcitedefaultendpunct}{\mcitedefaultseppunct}\relax
\EndOfBibitem
\bibitem[Fern{\'a}ndez-Delgado \latin{et~al.}(2014)Fern{\'a}ndez-Delgado,
  Cernadas, Barro, and Amorim]{fernandez2014we}
Fern{\'a}ndez-Delgado,~M.; Cernadas,~E.; Barro,~S.; Amorim,~D. Do We Need
  Hundreds of Classifiers to Solve Real World Classification Problems?
  \emph{Journal of Machine Learning Research} \textbf{2014}, \emph{15},
  3133--3181\relax
\mciteBstWouldAddEndPuncttrue
\mciteSetBstMidEndSepPunct{\mcitedefaultmidpunct}
{\mcitedefaultendpunct}{\mcitedefaultseppunct}\relax
\EndOfBibitem
\bibitem[Cort{\'e}s-Ciriano and Bender(2018)Cort{\'e}s-Ciriano, and
  Bender]{cortes2018deep}
Cort{\'e}s-Ciriano,~I.; Bender,~A. Deep confidence: a computationally efficient
  framework for calculating reliable prediction errors for deep neural
  networks. \emph{J. Chem. Inf. Model.} \textbf{2018}, \emph{59},
  1269--1281\relax
\mciteBstWouldAddEndPuncttrue
\mciteSetBstMidEndSepPunct{\mcitedefaultmidpunct}
{\mcitedefaultendpunct}{\mcitedefaultseppunct}\relax
\EndOfBibitem
\bibitem[Kendall and Gal(2017)Kendall, and Gal]{kendall2017uncertainties}
Kendall,~A.; Gal,~Y. What uncertainties do we need in bayesian deep learning
  for computer vision? Advances in Neural Information Processing Systems. 2017;
  pp 5574--5584\relax
\mciteBstWouldAddEndPuncttrue
\mciteSetBstMidEndSepPunct{\mcitedefaultmidpunct}
{\mcitedefaultendpunct}{\mcitedefaultseppunct}\relax
\EndOfBibitem
\bibitem[Krizhevsky \latin{et~al.}(2012)Krizhevsky, Sutskever, and
  Hinton]{krizhevsky2012imagenet}
Krizhevsky,~A.; Sutskever,~I.; Hinton,~G.~E. Imagenet classification with deep
  convolutional neural networks. Advances in Neural Information Processing
  Systems. 2012; pp 1097--1105\relax
\mciteBstWouldAddEndPuncttrue
\mciteSetBstMidEndSepPunct{\mcitedefaultmidpunct}
{\mcitedefaultendpunct}{\mcitedefaultseppunct}\relax
\EndOfBibitem
\bibitem[LeCun \latin{et~al.}(2015)LeCun, Bengio, and Hinton]{lecun2015deep}
LeCun,~Y.; Bengio,~Y.; Hinton,~G. Deep learning. \emph{Nature} \textbf{2015},
  \emph{521}, 436--444\relax
\mciteBstWouldAddEndPuncttrue
\mciteSetBstMidEndSepPunct{\mcitedefaultmidpunct}
{\mcitedefaultendpunct}{\mcitedefaultseppunct}\relax
\EndOfBibitem
\bibitem[Schmidt \latin{et~al.}(2019)Schmidt, Marques, Botti, and
  Marques]{schmidt2019recent}
Schmidt,~J.; Marques,~M.~R.; Botti,~S.; Marques,~M.~A. Recent advances and
  applications of machine learning in solid-state materials science. \emph{npj
  Comput. Mater.} \textbf{2019}, \emph{5}, 1--36\relax
\mciteBstWouldAddEndPuncttrue
\mciteSetBstMidEndSepPunct{\mcitedefaultmidpunct}
{\mcitedefaultendpunct}{\mcitedefaultseppunct}\relax
\EndOfBibitem
\bibitem[Domhan \latin{et~al.}(2015)Domhan, Springenberg, and
  Hutter]{domhan2015speeding}
Domhan,~T.; Springenberg,~J.~T.; Hutter,~F. Speeding up automatic
  hyperparameter optimization of deep neural networks by extrapolation of
  learning curves. International Joint Conference on Artificial Intelligence.
  2015\relax
\mciteBstWouldAddEndPuncttrue
\mciteSetBstMidEndSepPunct{\mcitedefaultmidpunct}
{\mcitedefaultendpunct}{\mcitedefaultseppunct}\relax
\EndOfBibitem
\bibitem[Kolachina \latin{et~al.}(2012)Kolachina, Cancedda, Dymetman, and
  Venkatapathy]{kolachina2012prediction}
Kolachina,~P.; Cancedda,~N.; Dymetman,~M.; Venkatapathy,~S. Prediction of
  learning curves in machine translation. The 50th Annual Meeting of the
  Association for Computational Linguistics. 2012; pp 22--30\relax
\mciteBstWouldAddEndPuncttrue
\mciteSetBstMidEndSepPunct{\mcitedefaultmidpunct}
{\mcitedefaultendpunct}{\mcitedefaultseppunct}\relax
\EndOfBibitem
\bibitem[Chow(1957)]{chow1957optimum}
Chow,~C.-K. An optimum character recognition system using decision functions.
  \emph{IRE Transactions on Electronic Computers} \textbf{1957}, 247--254\relax
\mciteBstWouldAddEndPuncttrue
\mciteSetBstMidEndSepPunct{\mcitedefaultmidpunct}
{\mcitedefaultendpunct}{\mcitedefaultseppunct}\relax
\EndOfBibitem
\bibitem[El-Yaniv and Wiener(2010)El-Yaniv, and Wiener]{el2010foundations}
El-Yaniv,~R.; Wiener,~Y. On the foundations of noise-free selective
  classification. \emph{Journal of Machine Learning Research} \textbf{2010},
  \emph{11}, 1605--1641\relax
\mciteBstWouldAddEndPuncttrue
\mciteSetBstMidEndSepPunct{\mcitedefaultmidpunct}
{\mcitedefaultendpunct}{\mcitedefaultseppunct}\relax
\EndOfBibitem
\bibitem[Geifman and El-Yaniv(2017)Geifman, and El-Yaniv]{geifman2017selective}
Geifman,~Y.; El-Yaniv,~R. Selective classification for deep neural networks.
  Advances in Neural Information Processing Systems. 2017; pp 4878--4887\relax
\mciteBstWouldAddEndPuncttrue
\mciteSetBstMidEndSepPunct{\mcitedefaultmidpunct}
{\mcitedefaultendpunct}{\mcitedefaultseppunct}\relax
\EndOfBibitem
\bibitem[Hendrycks \latin{et~al.}(2019)Hendrycks, Mazeika, and
  Dietterich]{hendrycks2018deep}
Hendrycks,~D.; Mazeika,~M.; Dietterich,~T. Deep Anomaly Detection with Outlier
  Exposure. International Conference on Learning Representations. 2019\relax
\mciteBstWouldAddEndPuncttrue
\mciteSetBstMidEndSepPunct{\mcitedefaultmidpunct}
{\mcitedefaultendpunct}{\mcitedefaultseppunct}\relax
\EndOfBibitem
\bibitem[Zweig and Campbell(1993)Zweig, and Campbell]{zweig1993receiver}
Zweig,~M.~H.; Campbell,~G. Receiver-operating characteristic (ROC) plots: a
  fundamental evaluation tool in clinical medicine. \emph{Clinical Chemistry}
  \textbf{1993}, \emph{39}, 561--577\relax
\mciteBstWouldAddEndPuncttrue
\mciteSetBstMidEndSepPunct{\mcitedefaultmidpunct}
{\mcitedefaultendpunct}{\mcitedefaultseppunct}\relax
\EndOfBibitem
\bibitem[Sugiyama \latin{et~al.}(2007)Sugiyama, Krauledat, and
  M{\~A}{\v{z}}ller]{sugiyama2007covariate}
Sugiyama,~M.; Krauledat,~M.; M{\~A}{\v{z}}ller,~K.-R. Covariate shift
  adaptation by importance weighted cross validation. \emph{Journal of Machine
  Learning Research} \textbf{2007}, \emph{8}, 985--1005\relax
\mciteBstWouldAddEndPuncttrue
\mciteSetBstMidEndSepPunct{\mcitedefaultmidpunct}
{\mcitedefaultendpunct}{\mcitedefaultseppunct}\relax
\EndOfBibitem
\bibitem[Krizhevsky(2009)]{krizhevsky2009learning}
Krizhevsky,~A. \emph{Learning Multiple Layers of Features from Tiny Images};
  University of Toronto, 2009\relax
\mciteBstWouldAddEndPuncttrue
\mciteSetBstMidEndSepPunct{\mcitedefaultmidpunct}
{\mcitedefaultendpunct}{\mcitedefaultseppunct}\relax
\EndOfBibitem
\end{mcitethebibliography}

\begin{tocentry}

\includegraphics[width=.9\textwidth]{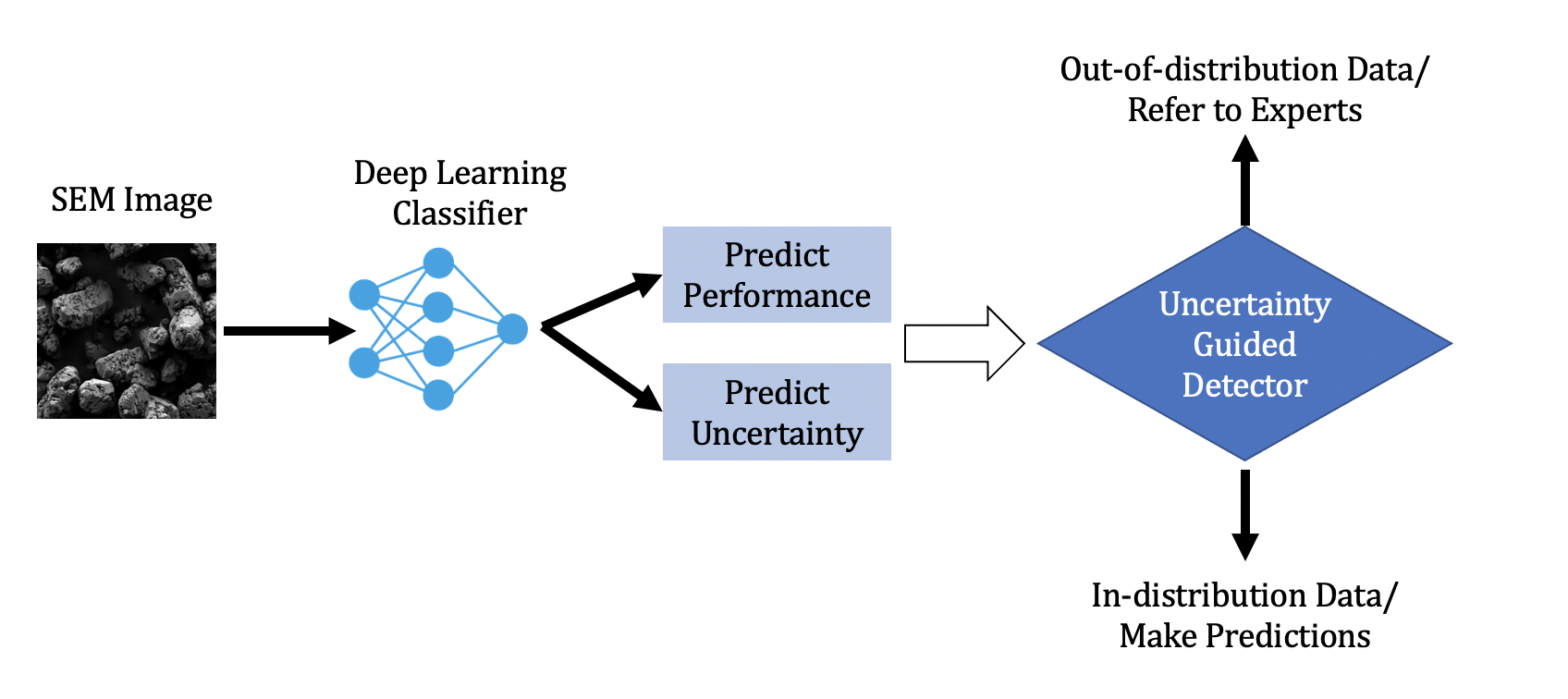}

\end{tocentry}

\end{document}